\documentclass[aps,prx,twocolumn,superscriptaddress, nofootinbib]{revtex4-2}
\usepackage[usenames,dvipsnames]{xcolor}
\usepackage{amssymb}
\usepackage{graphicx}
\usepackage{amsmath}
\usepackage{bbm}
\usepackage[bookmarks=true,colorlinks,citecolor=blue,urlcolor=blue]{hyperref}
\usepackage{braket}
\usepackage{babel}
\usepackage{blindtext}
\usepackage[compat=1.1.0]{tikz-feynman}
\usepackage[normalem]{ulem}
\usepackage{physics}
\usepackage{mathrsfs}
\usepackage{footmisc}
\usepackage{booktabs}
\usepackage{cancel}

\usepackage{mathtools}

\usepackage{dsfont}

\definecolor{mypurp}{rgb}{0.35, 0, 0.7}

\usepackage{amsthm}
\usepackage{txfonts}

\usepackage{soul}

\theoremstyle{definition}

\newcommand{\tGH}
{\text{FS}}
\newcommand{\tTC}
{\text{TC}}
\newcommand{\tcl}
{\text{Cl}}
\newcommand{\ttri}
{\text{tri}}
\newcommand{\tCX}
{\text{CX}}
\newcommand{\tSPT}
{\text{SPT}}
\newcommand{\ttop}
{\text{topo}}

\begin{document}

\title{Entanglement Properties of Gauge Theories from Higher-Form Symmetries}

\newcommand{\TUM}{\affiliation{Technical University of Munich, TUM School of Natural Sciences, Physics Department, 85748 Garching, Germany}}
\newcommand{\MCQST}{\affiliation{Munich Center for Quantum Science and Technology (MCQST), Schellingstr. 4, 80799 M{\"u}nchen, Germany}}
\newcommand{\Stanford}{\affiliation{Department of Physics, Stanford University, Stanford, California 94305, USA}}

\author{Wen-Tao Xu} \TUM \MCQST
\author{Tibor Rakovszky} \Stanford
\author{Michael Knap}  \TUM \MCQST
\author{Frank Pollmann}  \TUM \MCQST

\begin{abstract}

We explore the relationship between higher-form symmetries and entanglement properties in lattice gauge theories with discrete gauge groups, which can exhibit both topologically ordered phases and higher-form symmetry-protected topological (SPT) phases. Our study centers on a generalization of the Fradkin-Shenker model describing $\mathbb{Z}_2$ lattice gauge theory with matter, where the Gauss law constraint can be either emergent or exact. The phase diagram includes a topologically ordered deconfined phase and a non-trivial SPT phase protected by a 1-form and a 0-form symmetry, among others. We obtain the following key findings: First, the entanglement properties of the model depend on whether the 1-form symmetries and the Gauss law constraint are exact or emergent. For the emergent Gauss law, the entanglement spectrum (ES) of the non-trivial SPT phase exhibits degeneracies, which are robust at low energies against weak perturbations that explicitly break the exact 1-form symmetry. When the Gauss law and the 1-form symmetry are both exact, the ES degeneracy is extensive. This extensive degeneracy turns out to be fragile and can be removed completely by infinitesimal perturbations that explicitly break the exact 1-form symmetry while keeping the Gauss law exact. Second, we consider the ES in the topologically ordered phase where 1-form symmetries are spontaneously broken. In contrast to the ES of the non-trivial SPT phase, we find that spontaneous higher-form symmetry breaking removes “half” of the ES levels, leading to a non-degenerate ES in the topologically ordered phase in general. Third, we derive a connection between spontaneous higher-form symmetry breaking and the topological entanglement entropy (TEE). Using this relation, we investigate the entanglement entropy (EE) that can be distilled in the deconfined phase of the original Fradkin-Shenker model using gauge-invariant measurements. We show that the TEE is robust against the measurement when the 1-form symmetry is emergent  but it is fragile when the 1-form symmetry is exact. Our results demonstrate the advantage of higher-form symmetries for understanding entanglement properties of gauge theories.

\end{abstract}
\maketitle

\tableofcontents

\section{Introduction}

Much effort in condensed matter physics has been devoted to understanding the kinds of ordered phases that can appear in quantum many-body systems at low temperatures. A particularly detailed understanding exists for zero temperature properties of gapped local Hamiltonians, where phases 
roughly fall into three main categories: spontaneous symmetry breaking phases, intrinsic topological phases~\cite{LU_LRE_2010} and symmetry protected topological (SPT) phases~\cite{Pollmann_2011,chen2013symmetry}.
For intrinsic topological phases, a very fruitful approach has been to utilize (discrete) gauge theories, which provide an effective low-energy description of such phases~\cite{kitaev_2002,savary2016quantum}.
Recent work has brought to light a new perspective on these phases in terms of so-called \emph{higher-form symmetries}~\cite{Zohar_2004,NUSSINOV_2009,High_form_Kapistin_2015,McGreevy_2023}, which appear naturally in gauge theories and generalize the notion of usual ($0$-form) symmetries. While $0$-form symmetries act on the entire system, higher-form symmetries are associated with lower dimensional sub-manifolds.
Topological order can then be re-interpreted as a spontaneous breaking of such generalized symmetries, establishing an analogy with conventional spontaneous symmetry breaking phases.

\begin{figure}
    \centering
    \includegraphics{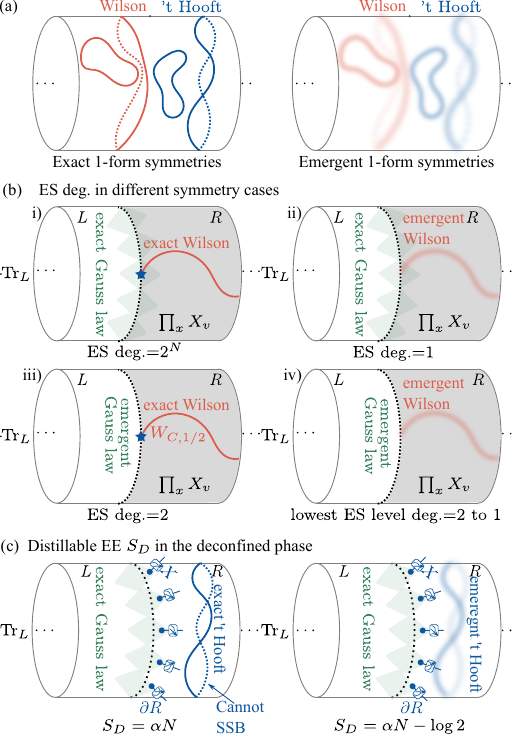}
    \caption{\textbf{1-form symmetries and main results.} We consider systems on an infinitely long cylinder, and a bipartition of the system into the left ($L$) and right ($R$) parts. $\partial R$ is the boundary of the $R$ part, $N$ is the circumference of the cylinder. (a) The 1-form Wilson and 't Hooft symmetries on contractible and non-contractible loops. When they are emergent symmetries, they correspond to ``broadened'' loop operators on the lattice.
    (b) ES degeneracy (indicated below the cylinder) 
    for four distinct situations: i) exact Wilson loop symmetry and exact Gauss law; ii) emergent Wilson loop symmetry and exact Gauss law; iii) exact Wilson loop symmetry and emergent Gauss law; vi) emergent Wilson loop symmetry and emergent Gauss law. 
    (c) In the deconfined phase of the Fradkin-Shenker model, the distillable EE satisfies an area law, $S_D=\alpha N$, without TEE when the 1-form 't Hooft loop symmetry is an exact symmetry. By contrast, when the 1-form 't Hooft loop symmetry is emergent, the TEE is present. 
   }     \label{Figure_0}
\end{figure}

An \emph{exact} higher-form symmetry corresponds to operators that commute with the microscopic Hamiltonian. A paradigmatic example of exact 1-form symmetries is given by the Wilson and 't Hooft loop operators in the toric code (TC) model in 2 dimensions, as shown in Fig.~\ref{Figure_0}a; both commute with the Hamiltonian. In the ground state manifold (which is 4-fold degenerate on a torus), these symmetries are spontaneously broken: the ground states transform non-trivially under the symmetry operators associated to non-contractible loops. More generally, higher-form symmetries can lead to new phases of matter, even when they are not spontaneously broken. In particular, they can protect non-trivial SPT phases~\cite{yoshida2016topological,Tsui_Wen_2020,pivot_2023,Ruben_Higgs_SPT,Higgs=SPT_second}. Once again, gauge theories provide a natural context where such phases can appear: it was pointed out recently~\cite{Ruben_Gauging_kitaev_2021,Ruben_Higgs_SPT,Higgs=SPT_second} that, within a certain parameter regime, the Higgs phase of a gauge theory can be interpreted as an SPT phase protected by a combination of 0-form and higher-form symmetries. 

An important difference between 0-form and higher-form symmetries is the enhanced robustness of the latter: even when the symmetry is explicitly broken at the microscopic level, it can still exist as an \emph{emergent symmetry} at low energies. This is well-known in the case of topological order, which is robust to any weak local perturbation. For example, in a perturbed toric code, one can always find broadened Wilson and `t Hooft loops that satisfy the same algebraic relations as the originals and serve as emergent symmetry operators~\cite{Hastings_and_Wen_2005}. While, perhaps counter-intuitively, the notion of emergent higher-form symmetry is more difficult to make precise when it is \emph{not} spontaneously broken, general arguments can be made that some notion of emergent symmetry should still survive~\cite{iqbal2022mean,Wen_emergent_high_form_2023}. 
It has also been pointed out that one can define a notion of emergent 1-form symmetry in the classical statistical mechanics formulation of a $\mathbb{Z}_2$ gauge theory~\cite{Adam_Nahum_2021,serna2024worldsheet}.

A commonly used and fruitful approach to characterize topological phases is in terms of the \emph{entanglement} properties of their ground states~\cite{Quantum_info_Quantum_matter}. Intrinsic topological order, in particular, is defined in terms of long-range entanglement~\cite{LU_LRE_2010} and can be diagnosed via the so-called topological entanglement entropy (TEE)~\cite{TEE_Kitaev_Preskill_2006, TEE_levin_Wen_2006,TEE_Renyi_wen_2009}. Furthermore, the entanglement spectrum (ES), i.e., the spectrum of the entanglement Hamiltonian (EH) defined as the negative logarithm of the reduced density matrix, can reflect boundary physics of a quantum system through the bulk-boundary correspondence~\cite{Li_Haldane_2008}, and detect non-trivial SPT phases of matter~\cite{Pollmann_2011} even in the absence of a physical boundary.

In this work, we study the interplay between higher-form symmetries and entanglement properties, in both topologically ordered and SPT phases. As a natural context where such symmetries arise, we consider a (2+1)-dimensional $\mathbb{Z}_2$ lattice gauge theory coupled to an Ising matter field~\cite{Kogut_LGH,Kogut_Susskind,TC_multi_critical_2010}, which is usually called the Fradkin-Shenker model. However, as we shall see, the Fradkin-Shenker model also has some non-generic entanglement properties for these phases, due to the presence of the exact Gauss law constraint. For this reason, we consider a generalized model (see
Sec.~\ref{TC_and_IGH}), where we also allow terms in the Hamiltonian that violate the Gauss law constraint, which is natural from the perspective of condensed matter systems where gauge theories are not ``fundamental'' but appear as emergent low-energy descriptions. Indeed, when such terms are small, the Gauss law can still be emergent at low energies, leading us to call the model an ``emergent'' $\mathbb{Z}_2$ lattice gauge theory.

For certain parameter regimes, the emergent $\mathbb{Z}_2$ lattice gauge theory exhibits exact 1-form symmetries, which can also survive as low-energy emergent symmetries when the exact symmetry is broken by sufficiently weak perturbations. We will be interested in the entanglement properties of the ground states in the various parts of the phase diagram and find that the entanglement properties of the model depends on whether the 1-form symmetries and the Gauss law constraint are exact or emergent. Numerically, we use two-dimensional tensor network states~\cite{maeshima:2001,verstraete2004}, which allow both for finding the ground states variationally or perturbatively and for efficiently extracting their entanglement properties.

First, in Sec.~\ref{sec:EH_ES_SPT} we investigate how the aforementioned 1-form SPT phase
manifests itself in the entanglement spectrum (ES) of the ground states. We are particularly interested in whether the robustness of physical edge modes to 1-form symmetry-breaking perturbations, observed in Ref.~\cite{Ruben_Higgs_SPT}, corresponds to a robust ES degeneracy. 
As summarized in Fig.~\ref{Figure_0}b, we find the following. i) When both the 1-form Wilson loop symmetry and the Gauss law constraint are exact, the ES has an \emph{extensive} $2^N$-fold degeneracy (with $N$ being the length of the entanglement cut). ii) When the Gauss law constraint is exact, this extensive degeneracy is lifted \emph{completely} by infinitesimal perturbations that explicitly break the exact 1-form Wilson loop symmetry 
yielding an ES with \emph{no} necessary degeneracy remaining.
This result is in contrast with the behavior of the physical boundary spectrum observed in Ref.~\cite{Ruben_Higgs_SPT}. 
iii) When the 1-form symmetry is exact, the ES degeneracy reduces from $2^N$-fold to $2$-fold by infinitesimal perturbations that explicitly break the exact Gauss law constraint (such that the Gauss law constraint becomes emergent). We show that this $2$-fold ES degeneracy cannot be removed without leaving the bulk SPT phase. iv) 
The remaining $2$-fold ES degeneracy becomes robust at low energies against weak perturbations that explicitly break the exact 1-form symmetry. 
Thus, in the phase diagram of the emergent $\mathbb{Z}_2$ lattice gauge theory, we identify the region with at least 2-fold exact ES degeneracy [situation i) and iii)] as the non-trivial 1-form SPT phase, and in the vicinity of the non-trivial SPT phase [situation iv)], the lowest ES level could exhibit 2-fold degeneracy in the thermodynamic limit. However, in the Higgs regime with the emergent 1-form Wilson loop symmetry and exact Gauss law [situation ii)], the ES is generally non-degenerate. 

Second, in Sec.~\ref{Sec:EH_ES_TC} we consider the ES in the topologically ordered phase where 1-form symmetries are spontaneously broken. We find that the EH near the fixed point of this topologically ordered phase has similar terms to the one near the fixed-point limit of the non-trivial 1-form SPT phase. However, the EH in the topologically ordered phase also contains a non-local projector which removes half of the spectrum, leading to a non-degenerate ES in the topologically ordered phase in general. 

Third, in Sec.~\ref{distillable_TEE_IGH}, we focus on the EE of the topologically ordered (deconfined) phase. We first show that the
spontaneous 1-form symmetry breaking is essential for TEE. We then consider the gauge theory being fundamental, with the exact Gauss law constraint being an intrinsic property of the physical Hilbert space.
As a result, the Hilbert space does not admit a decomposition into a tensor product of local degrees of freedom, and the usual EE defined through the Schmidt decomposition of wavefunctions becomes unsuitable. Instead, one can define it through the notion of \emph{entanglement distillation}\footnote{Entanglement distillation refers to the process of transforming $n$ copies of a quantum state into $m$ bell pairs via LOCC operations. The distillable EE 
of a quantum state $\rho$ is the maximum rate at which one can asymptotically distill maximally entangled qubit pairs from many copies of $\rho$; specifically, the distillable EE is $\max(\lim_{n\rightarrow\infty} \frac{m}{n})$.}  from quantum information theory
based on local operations and classical communication (LOCC)\footnote{General local quantum operations with classical communication (LOCC) in a bipartite system with Hilbert
spaces $\mathcal{H}_A$ and $\mathcal{H}_B$ are sequences of protocols in which one party, say, Alice, performs a local quantum operation who transmits the label reflecting the outcome to
Bob, who then again perform a local quantum operation in general dependent on the outcome of the form $\rho\rightarrow \sum_{j=1}^{J}\sum_{k=1}^{K_j}(A_j\otimes B_k^{(j)})\rho (A_j\otimes B_k^{(j)})^{\dagger}$, where trace preservation requires $\sum_{j=1}^{J}A^{\dagger}_jA_j=\sum_{k=1}^{K_s}B^{\dagger}_kB_k=\mathbbm{1}$}.
In particular, one can define the entanglement of a fundamental gauge theory in terms of the 
entanglement distillation using only those LOCC operations that satisfy the exact Gauss law constraint~\cite{Hamma_2005,PRD_decompose_EE_2012,Distillable_entanglement_2016,Gauge_ent_1,Gauge_ent_2,Gauge_ent_3,Gauge_ent_4,Gauge_ent_5}.
We uncover a subtle interplay between the behavior of the distillable EE and the presence of higher-form symmetries. As shown in Fig.~\ref{Figure_0}c, we find that in a generic point of the deconfined phase, where the 1-form 't Hooft loop symmetry is only emergent, the distillable EE satisfies the area law minus the TEE. However, when the 1-form 't Hooft loop symmetry is \emph{exact}, the TEE is \emph{absent}.
The reason is that the entanglement distillation involves gauge-invarinat measurement (see Fig.~\ref{Figure_0}c), and the TEE is robust against the measurement when the 1-form 't Hooft loop symmetry is emergent  but it is fragile when the 1-form 't Hooft loop symmetry is exact.

Finally, discussion and outlook are presented in Sec.~\ref{discussion_and_outlook}. Technical details can be found in the Appendices.

\section{Emergent $\mathbb{Z}_2$ lattice gauge theory and symmetry transformations}\label{TC_and_IGH}

In this section, we first introduce an emergent $\mathbb{Z}_2$ lattice gauge theory, which is our standard model for studying the interplay between 1-form symmetries, the Gauss law constraint, and entanglement properties. We discuss how in various limits, our model reduces to the TC model in a field~\cite{kitaev_2002,Youjin_TC_phase_diagram_2011,TC_phase_diagram_expansion_2009}, the Fradkin-Shenker model~\cite{Kogut_LGH,Fradkin_1979,Kogut_Susskind,TC_multi_critical_2010}, as well as a Lieb-lattice cluster model in a field that realizes a non-trivial SPT phase protected by both a 1-form symmetry and a 0-form symmetry~\cite{yoshida2016topological}. We also discuss the isometry transformation (or equivalently fixing to unitary gauge) that relates the TC model in a field and the Fradkin-Shenker model, and how it leads to a non-trivial quantum channel transforming between their respective reduced density matrices.

\begin{figure}[h!]
    \centering
\includegraphics{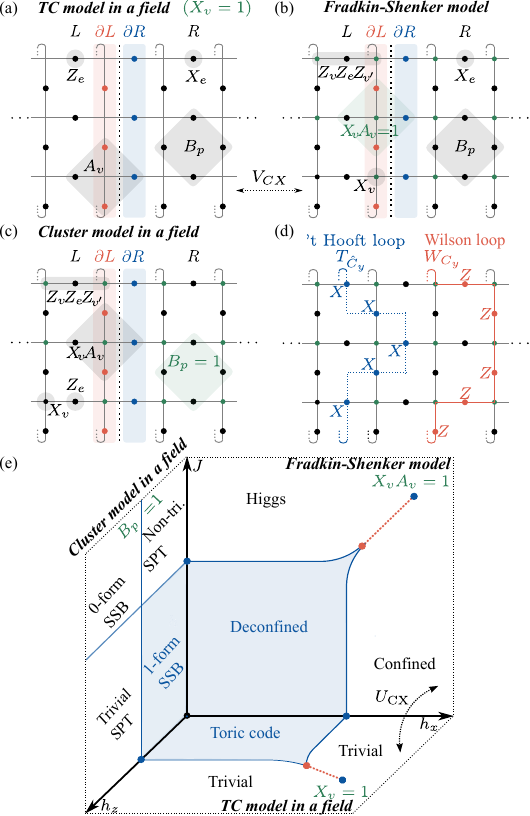}
\caption{\textbf{Models, entanglement bipartition, and phase diagram.} We consider
systems on an infinitely long cylinder. The entanglement bipartition cuts the system into the left (right) part $L$ ($R$),
where $\partial L\in L$ ($\partial R\in R$) are the boundaries of $L$ ($R$). (a) The TC model in a field consists of
qubits on the edges of a square lattice. $A_v=\prod_{e\in v} X_e$ ($B_p=\prod_{e\in p} Z_e$)
is defined by four $X$ ($Z$) operators belonging to a vertex (plaquette). 
(b) The lattice
of the Fradkin-Shenker model and the Hamiltonian terms. The qubits on edges
(vertices) are the $\mathbb{Z}_2$ gauge (matter) fields. The exact Gauss law constraint
is $X_vA_v=1,\forall v$. (c) The cluster model in a field on a Lieb lattice and the Hamiltonian terms. There is the exact 1-form Wilson loop symmetry, and the ground state satisfies $B_p=1,\forall p$. (d) The non-contractible Wilson and 't Hooft loop on a cylinder. (e) The phase diagram of the emergent $\mathbb{Z}_2$ lattice gauge theory, which contains the TC model in a field, the Fradkin-Shenker model, and the cluster model in a field. ``Toric code'', ``Deconfined'', and ``1-form SSB'' denote the same $\mathbb{Z}_2$ topologically ordered phase whose fixed point is at $(h_x,h_z,J)=(0,0,0)$. The fixed point of the non-trivial SPT phase is at $(h_x,h_z,J)=(0,0,+\infty)$. 
}\label{Entanglement_cut}
\end{figure}

The Hilbert space of our emergent $\mathbb{Z}_2$ lattice gauge theory is defined by qubits on edges and vertices of a square lattice. We label the edges, vertices, and plaquettes of the lattice as $e$, $v$, and $p$, respectively. The Hamiltonian reads~\cite{Ruben_Higgs_SPT}:
\begin{align}\label{EGT}
H(h_x,h_z,J)=&-\sum_v X_v-\sum_p B_p-J\sum_{\langle vev'\rangle} Z_vZ_eZ_{v'}-h_x\sum_eX_e\notag\\
&-h_z\sum_eZ_e-\sum_v X_vA_v,
\end{align}
where $X_e$ ($X_v$) and $Z_e$ ($Z_v$) are Pauli matrices on edges (vertices), $A_v=\prod_{e\in v}X_e$ and $B_p=\prod_{e\in p} Z_e$ are vertex and plaquette operators, and $\langle vev'\rangle$ denotes nearest neighbouring two vertices $v$ and $v^{\prime}$ and the edge $e$ between them, see Figs.~\ref{Entanglement_cut}a-c.

The first four terms in Eq.~\eqref{EGT} correspond to the usual Hamiltonian of the Fradkin-Shenker model. However, unlike the usual case, we are not enforcing the exact Gauss law, $X_v A_v = +1$ as a hard constraint on the physically allowed configurations, but implement it energetically. Thus, we allow for arbitrary configurations of the combined matter (vertex) and gauge (edge) degrees of freedom and the Hamiltonian itself contains a term $\sum_e Z_e$ which fails to commute with $X_vA_v$. 
At $h_z = 0$, the Hamiltonian becomes gauge-invariant, and imposes the exact Gauss law at low energies to recover the conventional $\mathbb{Z}_2$ gauge theory with the Ising matter field, along with its usual Fradkin-Shenker phase diagram~\cite{Fradkin_1979}; see below. Introducing $h_z$ allows us to probe the consequences of breaking the exact Gauss law. While all the phases of the Fradkin-Shenker model survive at small values of $h_z$ (hence justifying the name ``emergent'' gauge theory), 
we will see that the gauge theory at $h_z=0$ exhibits non-generic properties of the entanglement, particularly in the Higgs phase.

A crucial aspect of our analysis will be the symmetries of the model. First of all, for any choice of the couplings, the Hamiltonian in Eq.~\eqref{EGT} commutes with the global 0-form Ising symmetry $\prod_{v} X_v$. At the fixed point ($h_x=h_z=J=0$) of the topologically ordered phase, there is also a pair of exact 1-form symmetries, generated by the Wilson loops $W_C$ and the't Hooft loops $T_{\hat{C}}$ (Fig.~\ref{Entanglement_cut}d). They are defined as
\begin{equation}\label{def_of_loops}
   W_C=\prod_{e\in C}Z_e,  \quad T_{\hat{C}}=\prod_{e\in \hat{C}}X_e,
\end{equation}
where $C$ and  $\hat{C}$ are closed loops on the primal lattice and dual lattice, respectively. For arbitrary choices of $C$ and $\hat{C}$, we have $[W_C,H(0,0,0)]=[T_{\hat{C}},H(0,0,0)]=0$. Away from this fixed point, $W_C$ and $T_{\hat{C}}$ are generally no longer exact symmetries of the model. However, one of the 1-form symmetries survives for a certain choice of the parameters: for $h_x = 0$, the 1-form Wilson loop symmetry commutes
with $H(0, h_z,J): [W_C, H(0,h_z,J)] = 0,\forall C$, and for $h_z=J=0$, the
1-form ’t Hooft loop symmetry commutes with $H(h_x,0, 0)$: $[T_{\hat{C}}, H(h_x,0,0)] = 0,\forall \hat{C}$.
Crucially, even when the exact 1-form symmetries are weakly broken by sufficiently small values of the couplings, higher-form symmetries can still emerge at low energies, capturing the robustness of
topological order.  Concretely, both of the 1-form symmetries
are emergent at low energies in the $\mathbb{Z}_2$ topologically ordered phase shown in Fig.~\ref{Entanglement_cut}e, corresponding to ``broadened'' loop operators (illustrated in Fig.~\ref{Figure_0}a); an example is shown in Appendix~\ref{App:explicit_form_of_emergent_1-form_sym}. These broadened loop operators satisfy the same algebra as their unbroadened versions, implying topologically degenerate ground states. In
this formalism, topological order can be interpreted as a spontaneous breaking of such emergent symmetries, drawing analogies to conventional symmetry breaking phases~\cite{High_form_Kapistin_2015,High_form_wen_2019,McGreevy_2023,Wen_emergent_high_form_2023}. Higher form symmetries also help us to understand the topologically trivial phase. In particular, the trivial phase can be separated into regimes with an emergent 1-form symmetry that is not spontaneously broken and without the emergent 1-form symmetry~\cite{Adam_Nahum_2021,Wen_emergent_high_form_2023}. 

As mentioned before, when $h_x = 0$, the Wilson loop symmetry remains an \emph{exact}  1-form symmetry. Of particular interest to us is the point $h_x = h_z = 0$ and $J\to +\infty$, where the model reduces to the so-called cluster model on the Lieb lattice (which is obtained from the square lattice by adding a site in the middle of every bond). This is known to be an example of a non-trivial SPT phase, protected by the combination of the 0-form Ising symmetry and the 1-form Wilson loop symmetry~\cite{yoshida2016topological,Ruben_Higgs_SPT}. Since making either $h_z$ or $1/J$ finite does not break these symmetries, this phase survives up to a finite value of $h_z$ and $1/J$. Adding the $h_x$ field, however, explicitly breaks the exact Wilson loop symmetry. The fate of the non-trivial SPT phase in the presence of a finite $h_x$ field will be one of the central questions we investigate below.

As the previous discussion already suggests, it will be useful to consider the three extreme limits, with either $J=0$, $h_z=0$ or $h_x=0$, corresponding to two-dimensional planes of the 3-parameter phase diagram (see Fig.~\ref{Entanglement_cut}e). In the first case, $J=0$, the matter and gauge fields decouple; the former becomes a product state, and the latter forms a TC model in a field. The second case $h_z=0$ corresponds to the Fradkin-Shenker model. The third one $h_x=0$ is equivalent to the the cluster model in a field. We now discuss these three planes in more detail.

\subsection{Tilted field toric code model}

When $J=0$, $X_v$ is a conserved quantity of the Hamiltonian:  $[H(h_x,h_z,0),X_v]=0,\forall v$, and in the Hilbert space satisfying $X_v=1,\forall v$, the emergent $\mathbb{Z}_2$ lattice gauge theory reduces to the TC model in a field~\cite{kitaev_2002,Youjin_TC_phase_diagram_2011,TC_phase_diagram_expansion_2009}:
 \begin{equation}\label{TC_Hamiltonian}
H_{\text{TC}}=-\sum_{v}A_v-\sum_{p}B_p-h_x\sum_{e}X_e-h_z\sum_{e} Z_e,
\end{equation}
see Fig.~\ref{Entanglement_cut}a. Notice that the TC model in a field only involves qubits on edges since the vertex qubits are fixed to $\ket{+}_v$ by $X_v=1$.
The phase diagram of the TC model in a field~\cite{Youjin_TC_phase_diagram_2011, TC_phase_diagram_expansion_2009}, exhibiting a toric code phase with the $\mathbb{Z}_2$ topological order and a trivial phase, is shown in the $(h_x.h_z)$ plane of Fig.~\ref{Entanglement_cut}e. The continuous transition from the toric code phase to the trivial phase belongs to the $(2+1)$D Ising universality class.

\subsection{Fradkin-Shenker model}
When $h_z=0$, the term $X_vA_v$ is a conserved quantity of the Hamiltonian: $[H(h_x,0,J),X_vA_v]=0,\forall v$. In the Hilbert space constrained by the exact Gauss law, $X_vA_v=1,\forall v$, the emergent $\mathbb{Z}_2$ lattice gauge theory reduces to the Fradkin-Shenker model~\cite{Kogut_LGH,Kogut_Susskind,TC_multi_critical_2010}:
\begin{equation}\label{GH_Hamiltonian}
H_{\tGH}=-\sum_{v}X_{v}-\sum_{p}B_p-h_x\sum_{e}X_e-J\sum_{\langle vev' \rangle} Z_{v}Z_eZ_{v'},
\end{equation}
see Fig.~\ref{Entanglement_cut}b.  And $J$ ($h_x$) can be regarded as the strength of the gauge-matter coupling (gauge fluctuations).

The phase diagram of the Fradkin-Shenker model is shown in the $(h_x,J)$ plane of Fig.~\ref{Entanglement_cut}e.
The deconfined phase of the Fradkin-Shenker model has $\mathbb{Z}_2$ topological order.
At sufficiently large gauge-matter coupling $J$ (gauge fluctuation $h_x$), there is a phase transition to the Higgs (confining) regime in which $\mathbb{Z}_2$ charges condense (are confined). For a sufficiently large $h_x$ field, there is a phase transition to the confining regime in which $\mathbb{Z}_2$ charges are confined.
With the periodic boundary condition, the Higgs and the confined regimes are adiabatically connected, and thus, they belong to the same phase~\cite{Fradkin_1979}.

When the gauge-matter coupling is set to $J=0$ (and keeping $h_z=0$), the Fradkin-Shenker model reduces to a pure $\mathbb{Z}_2$ lattice gauge theory with the exact 1-form 't Hooft loop symmetry $T_{\hat{C}}$, and the gauge and matter fields are not entangled, i.e.,  $\ket{\Psi_{\tGH}}=\ket{\Psi_{\tTC}}\otimes\prod_v\ket{+}_v$. Thus the pure $\mathbb{Z}_2$ lattice gauge theory and the TC model in a transverse field are the same model and their ground states $\ket{\Psi_{\tGH}}$ and $\ket{\Psi_{\tTC}}$ only differ by a product state.

For $h_x=0$, the Fradkin-Shenker model has the exact 1-form Wilson loop symmetry $W_C$. In the limit of $J \gg 1$, this symmetry is not broken. However, as shown in Ref.~\cite{Ruben_Higgs_SPT}, this regime realizes a non-trivial SPT phase protected by a 1-form $\mathbb{Z}_2$ symmetry $W_C$ and the 0-form Ising symmetry $\prod_vX_v$.
This SPT can be detected by a string order parameter in the bulk and a 2-fold degeneracy in the presence of symmetry-preserving boundary conditions. It was found in Ref.~\cite{Ruben_Higgs_SPT} that for appropriately chosen boundary terms, this physical boundary degeneracy also survives in the Higgs regime with a small gauge fluctuation $h_x$ (which breaks the exact 1-form Wilson loop symmetry) until a boundary phase transition occurs. Relatedly, it has been argued that the 1-form symmetry can survive as an emergent symmetry at small $h_x$ in the Higgs regime~\cite{Adam_Nahum_2021,Wen_emergent_high_form_2023}.

Ref. \cite{Ruben_Higgs_SPT} raised the question of whether the same kind of robust degeneracy is present in the ES, which could provide a bulk characterization of the SPT phase even when the 1-form symmetry is not exact. As we shall see, this is not the case: the ES of the Fradkin-Shenker model has no degeneracies when $h_x > 0$. However, we will see later that there exists a 2-fold ES degeneracy in
the emergent $\mathbb{Z}_2$ lattice gauge theory, in a regime where the Gauss law constraint is emergent.

\subsection{Cluster model in a field}\label{sec:cluster_model}

For $h_x=0$, the emergent $\mathbb{Z}_2$ lattice gauge theory obeys the exact 1-form Wilson loop symmetry, generated by the plaquette terms $B_p$. As such, at low energies, we can fix $B_p = +1$, which reduces the Hamiltonian to that of the Lieb lattice cluster model in a field:
\begin{equation}\label{Ham_cluster}
H_{\tcl}=-\sum_v X_vA_v-J\sum_{\langle vev' \rangle} Z_{v}Z_eZ_{v'}-\sum_v X_v-h_z\sum_e Z_e,
\end{equation}
see Fig.~\ref{Entanglement_cut}c. 
Note that the cluster model in a field breaks the exact Gauss law constraint when $h_z\neq 0$.

Let us consider some limiting cases of the cluster model in a field. First, when $h_z=J=0$, the qubits on the edges of the lattice form a fixed point TC ground state, which belongs to the $\mathbb{Z}_2$ topologically ordered phase with spontaneous 1-form symmetry breaking. Keeping $J=0$ but taking $h_z \to \infty$, $H_{\tcl}$ reduces to the fixed point Hamiltonian of the trivial SPT phase: $H_{\ttri}=-\sum_v X_v-h_z\sum_e Z_e$, whose ground state is a product state. On the other hand, when $J$ is very large and $h_z =0$, Eq.~\eqref{Ham_cluster} effectively reduces to the Lieb-lattice cluster model
\begin{equation}\label{Ham_SPT}
H_{\tSPT}=-\sum_v X_vA_v-J\sum_{\langle vev' \rangle} Z_{v}Z_eZ_{v'},
\end{equation}
which is a fixed point Hamiltonian of the non-trivial SPT phase protected by the exact 1-form Wilson loop symmetry and the 0-form Ising symmetry~\cite{yoshida2016topological,Ruben_Higgs_SPT}.
Finally, When $h_z,J=+\infty$, the qubits on vertices of the lattice form a ferromagnetic state, which spontaneously breaks the 0-form Ising symmetry.

These considerations show that the cluster model in a field in Eq.~\eqref{Ham_cluster} has at least four distinct phases. Actually, its phase diagram can be obtained exactly because $H_{\tcl}$ can be separated into two commuting parts, and each part can be mapped to the $(2+1)$D transverse field Ising model~\cite{Ruben_Higgs_SPT}. Consequently the phase boundaries are straight lines in $(h_z,J)$ plane of Fig.~\ref{Entanglement_cut}e.

\subsection{Quantum channel between reduced density matrices of the TC model in a field and the Fradkin-Shenker model}\label{subsec:Q-channel}

It is well known that the Fradkin-Shenker model is equivalent to the TC model in a field~\cite{TC_multi_critical_2010}. In particular, there exists a gauge choice (the so-called ``unitary gauge'') which eliminates the matter degrees of freedom. The spectrum of the Hamiltonian, and local physical properties, are unchanged under this transformation. However, we will be interested in entanglement properties, where the situation is less obvious. In fact, as we will show, the reduced density matrices in the two models are related to each other non-trivially by a quantum channel, which can lead to important differences between their ES.

Let us introduce the following transformation, which entangles matter and gauge fields via a series of local controlled-$X$ gates on all sets of adjacent vertices $v$ and edges $e$:
\begin{equation}
   U_{\tCX}=\prod_{\langle ve\rangle}\text{CX}_{ve},\quad \text{CX}_{ve}=\frac{1+Z_v}{2}\mathbbm{1}_e+\frac{1-Z_v}{2}X_e.
\end{equation}
$U_{\tCX}$ transforms Hamiltonian~\eqref{EGT} as
\begin{equation}
    U_{\tCX}H(h_x,h_z,J)U_{\tCX}^{\dagger}=H(h_x,J,h_z).
\end{equation}
This has several implications. First of all, in the cluster model in a field, at $h_x = 0$, this transformation acts as the ``SPT entangler'', which transforms between the trivial and non-trivial SPT phases: $H_{\ttri}=U_{\tCX}H_{\tSPT}U_{\tCX}^{\dagger}$, as well as the corresponding non-fixed point Hamiltonians. More generally, $U_{\tCX}$ relates the plane of the TC model in a field and the Fradkin-Shenker model plane in the phase diagram in Fig.~\ref{Entanglement_cut}e.
Since the matter qubits on vertices of the TC model in a field are fixed to $\ket{+}_v$ by $X_v=1$, we can discard them, $U_{\tCX}$ reduces to an isometry transformation $V_{\tCX}$ between the TC model in a field and the Fradkin-Shenker model~\cite{TC_multi_critical_2010}:
\begin{equation}
   H_{\tGH}=V_{\tCX}H_{\text{TC}}V_{\tCX}^{\dagger},\quad V_{\tCX}=U_{\tCX}\prod_{v}\ket{+}_v,\label{eq:iso}
\end{equation}
where $V_{\tCX}$ satisfies
 \begin{equation}
     V_{\tCX}^{\dagger}V_{\tCX}=\mathbbm{1},\quad V_{\tCX}V_{\tCX}^{\dagger}=\prod_v [(1+X_vA_v)/2].
 \end{equation}
From the perspective of a usual gauge theory, the effect of $V_{\tCX}^\dagger$ is equivalent to going into the unitary gauge~\cite{Ruben_Higgs_SPT}.
Since $V_{\tCX}$ is an isometry, which preserves the bulk energy spectrum, the Fradkin-Shenker model and the TC model in a field share the same phase diagram.

The ground states $\ket{\Psi_{\tGH}}$ of the Fradkin-Shenker model can thus be obtained from the ground states $\ket{\Psi_{\text{TC}}}$ of the TC model:
\begin{equation}\label{GS_transformation}
    \ket{\Psi_{\tGH}}=U_{\tCX}\left(\ket{\Psi_{\tTC}}\otimes \prod_v\ket{+}\right)=V_{\tCX}\ket{\Psi_{\tTC}}.
\end{equation}
Considering the bipartition shown in Figs.~\ref{Entanglement_cut}a and b and tracing all qubits in the $L$ part, the reduced density matrices of the TC model in a field and the Fradkin-Shenker model can be defined as $\rho_{\tTC}=\Tr_L\ket{\Psi_{\tTC}}\bra{\Psi_{\tTC}}$ and $\rho_{\tGH}=\Tr_L\ket{\Psi_{\tGH}}\bra{\Psi_{\tGH}}$. Since the ground states are related by Eq.~\eqref{GS_transformation}, a natural question is: what is the relation between their reduced density matrices?
\begin{figure}
    \centering
    \includegraphics{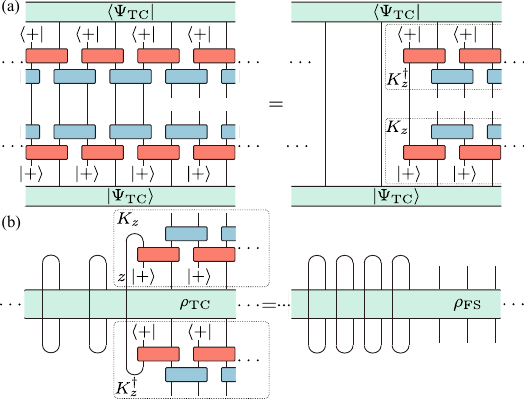}
    \caption{\textbf{Quantum channel relating reduced density matrices of the TC model in a field and Fradkin-Shenker model.} For clarity, we illustrate the quantum channel for reduced density matrices in 2 dimensions using 1-dimensional plots. Red and blue rectangles are CX gates in different directions (CX gates are not spatial inversion symmetric).  (a) Calculation of $\rho_{\tGH}$ from $\ket{\Psi_{\tTC}}$ and CX gates. (b) The right side of (a) is a quantum channel applied to the reduced density matrix $\rho_{\tTC}$ of the TC model in a field, which gives rise to the reduced density matrix $\rho_{\tGH}$ of the Fradkin-Shenker model. The dashed line boxes denote the Kraus operators.}
    \label{quantum_channel}
\end{figure}

As shown in Fig.~\ref{quantum_channel}a, from $\ket{\Psi_{\tTC}}$ and the CX gates, we find that the reduced density matrices $\rho_{\tTC}$ and $\rho_{\tGH}$ are related by a completely positive map $\mathscr{N}[\cdot]$:
\begin{equation}\label{eq:quantum_channel}
    \rho_{\tGH}=\mathscr{N}[\rho_{\tTC}]=\sum_{\pmb{z}} K_{\pmb{z}}\rho_{\tTC} K^{\dagger}_{\pmb{z}},
\end{equation}
where $\ket{\pmb{z}}=\bigotimes_{v\in\partial L}\ket{z_v}$, $\partial L$ is the boundary of the region $L$ and $z_v=0$ or $1$, and $K_{\pmb{z}}$ is the Kraus operator
\begin{equation}\label{Kraus_operator}
    K_{\pmb{z}}=\frac{1}{2^{N/2}}\bra{\pmb{z}}\prod_{\langle e\in R,v\in R\cup\partial L\rangle }\text{CX}_{v,e}
   \prod_{v\in R\cup\partial L}\ket{+}_v,
\end{equation}
 as shown in Fig.~\ref{quantum_channel}b.
The completely positive map $\mathscr{N}[\cdot]$ satisfies the trace-preserving condition $\sum_{\pmb{z}}K_{\pmb{z}}^{\dagger}K_{\pmb{z}}=\mathbbm{1}$,
making it a quantum channel.
Since the quantum channel does not necessarily preserve the spectrum of a density matrix, the ES of the Fradkin-Shenker model is generically different from that of the TC model in a field. We will study the implications of this below. 

\section{Entanglement spectrum of the non-trivial SPT phase and its vicinity}\label{sec:EH_ES_SPT}

The phase diagram of Hamiltonian~\eqref{EGT}, shown in Fig.~\ref{Entanglement_cut}e, broadly divides into two regions: the topologically ordered phase near the origin, and the outside region, without topological order. In this section, we focus on the latter regime. We return to the entanglement properties of the topologically ordered phase below in Sections~\ref{Sec:EH_ES_TC} and~\ref{distillable_TEE_IGH}. Here, we consider the entanglement spectrum (ES) of the non-trivial SPT phase protected by both the exact 1-form Wilson loop symmetry and the 0-form Ising symmetry, as well as in its vicinity where the 1-form Wilson loop symmetry becomes an emergent symmetry. To simplify our numerical approach, we choose lattice with cylinder geometry and take a bipartition of the system into two half-cylinders, separated by an entanglement cut far from either boundary. In the topologically trivial phases we consider here, although degenerate ground states could stem from boundaries, the ES is independent of which ground state we choose.

To calculate the ES numerically, we first express the ground states in terms of a projected entangled pair states (PEPS)~\cite{maeshima:2001,verstraete2004}, and then combine the quantum channel and the standard method~\cite{Cirac_2011} to calculating ES from the PEPS, as shown in Appendix~\ref{Appendix_iPEPS}. In principle, the ground states of the emergent $\mathbb{Z}_2$ gauge theory can be obtained by variationally optimizing the PEPS. However, since we only consider the ES near the fixed point of gapped phases, we can alternatively use the PEPS derived from a perturbative calculation. We find that in the parameter regime we consider, the ES of the variational PEPS and the perturbed PEPS match very well, as we shown in Appendix~\ref{Appendix_iPEPS}. Thus, we adopt the perturbed PEPS to calculated the ES in the following sections. 

\subsection{ES degeneracy in and near the non-trivial SPT phase}

We begin by considering the fixed-point of the non-trivial SPT phase, given by Hamiltonian $H_{\tSPT}$ in Eq.~\eqref{Ham_SPT}, where we find an extensive degeneracy of the ES. Motivated by the boundary phase transition in Ref.~\cite{Ruben_Higgs_SPT}, we then consider a path within the plane of the Fradkin-Shenker model in Fig.~\ref{Entanglement_cut}e, which explicitly breaks the exact 1-form Wilson loop symmetry and connects the Higgs regime to the confined regime. We find that the ES is non-degenerate everywhere along this path when the exact 1-form symmetry is explicit broken and the Gauss law constraint is exact, without any phase transition. Next, we turn to a different perturbation of the cluster state, which preserves the exact 1-form Wilson loop symmetry but removes the exact Gauss law constraint, and we find that this reduces the ES degeneracy from an extensive one to a 2-fold one, which is necessary within the non-trivial SPT phase. Finally, we consider a generic path in the volume between the Higgs regime and the non-trivial SPT phase (see Fig.~\ref{Entanglement_cut}e), where both the exact 1-form Wilson loop symmetry and exact Gauss law constraint are emergent, and we show that there is a robust regime where the 2-fold  degeneracy is present in the lowest ES level. 

\subsubsection{Explicitly breaking the exact 1-form Wilson loop symmetry}

Let us start by examining the cluster model given in Eq.~\eqref{Ham_cluster}, which is a fixed point Hamiltonian of the non-trivial SPT phase. The corresponding fixed-point wavefunction has the form
\begin{equation}\label{cluster_state}
    \ket{\tcl}=U_{\tCX}\ket{\ttri}=\prod_v\frac{1+X_vA_v}{\sqrt{2}}\prod_e\ket{0}_e\otimes\prod_v\ket{0}_v.
\end{equation}
Motivated by Ref.~\cite{Ruben_Higgs_SPT}, we consider a path $(\delta'=\sqrt{h_x^2+J^2}=+\infty,\theta'=\tan^{-1}(J/h_x))$, which interpolates between the Higgs regime and the confined regime within the plane of the Fradkin-Shenker model. To understand the entanglement properties along this path, it will be useful to compare the corresponding path in the plane in the TC model in a field shown in Fig.~\ref{Entanglement_cut}e, given by $(\delta=\sqrt{h_x^2+h_z^{2}}=+\infty,\theta=\tan^{-1}(h_z/h_x))$. The two paths $\delta=+\infty$ and $\delta'=+\infty$ are related by the isometry transformation introduced in Sec.~\ref{subsec:Q-channel}.

In the TC model in a field at $\delta=+\infty$, the ground state becomes a product state $\prod_e\ket{\theta}_e$ along the whole path, with $\ket{\theta}=\cos(\theta/2)\ket{+}+\sin(\theta/2)\ket{-}$, and the reduced density matrix is simply $\rho_{\tTC}=\prod_{e\in\partial R}\ket{\theta}_e\bra{\theta}_e$.
We can then apply the quantum channel $\mathscr{N}[\cdot]$ introduced in Sec.~\ref{subsec:Q-channel} to $\rho_{\tTC}$ to get the corresponding reduced density matrix $\rho_{\tGH}$ of the Fradkin-Shenker model. When doing so, we observe that only those CX gates that cross the entanglement cut contribute to entanglement; all remaining gates act within the un-traced part and thus do not change the ES:
\begin{equation}
    \rho_{\tGH}=\mathscr{N}[\rho_{\tTC}]\sim\Tr_v\prod_{\langle e\in\partial R,v\in\partial L\rangle } \text{CX}_{v,e}\ket{+}_v\ket{\theta}_e\bra{\theta}_e\bra{+}_v\text{CX}^{\dagger}_{v,e},
\end{equation}
with $N$ being the circumference of the cylinder. From the above equation we can derive the effective entanglement Hamiltonian (EH):%
\begin{equation}\label{EH_GH_trivial}
   H^{\text{eff}}_{E,\tGH}=-\sum_{i=1}^{N}\text{arctanh}(\cos \theta') X_i+N\log[\sin(\theta')/2],
\end{equation}
where $X_i$ acts on the effective entanglement degrees of freedom, i.e., virtual bonds of a tensor network state. From $ H^{\text{eff}}_{E,\tGH}$, we see that the ES has an extensive $2^N$-fold degeneracy at $\theta=\pi/2$ but has a unique ground state otherwise.

\begin{figure}
    \centering
    \includegraphics[scale=0.5]{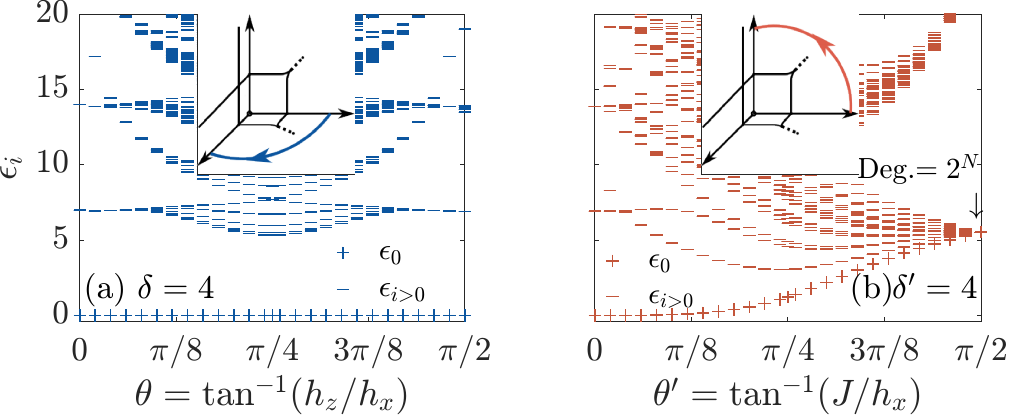}
    \caption{\textbf{ES of the TC model in a field and the Fradkin-Shenker model along $\delta=\delta'=4$.} ES $\{\epsilon_i\}$ ($\epsilon_0\ge\epsilon_1\ge \epsilon_2\ge\cdots$) for systems on an infinitely long cylinder with a finite circumference $N=8$, calculated from the PEPS constructed by perturbing from the $\delta=\delta'=+\infty$ limit, and the bond dimension of the boundary matrix product states (MPS) of the PEPS is $\chi=20$. 
    Insets show schematic paths of each panel in the phase diagram displayed in Fig.~\ref{Entanglement_cut}e.
    (a) ES in the trivial phase of the TC model in a field. (b) ES in the Higgs-confined phase of the Fradkin-Shenker model.}
    \label{fig:ES_Higgs_conf}
\end{figure}

To complement the above analytical argument, we calculate the ES numerically along a more generic path $\delta=\delta'=4$, from the PEPS constructed by perturbing from the limit $\delta=\delta'= +\infty$, see Appendix~\ref{Appendix_iPEPS}. The results agree qualitatively with the limit $\delta=\delta'= +\infty$. For the path $\delta=4$ within the trivial phase of the TC model in a field, shown in Fig.~\ref{fig:ES_Higgs_conf}a, the ES is non-degenerate along the entire path. The ES of the Fradkin-Shenker model along $\delta'=4$ is shown in Fig.~\ref{fig:ES_Higgs_conf}b. At $\delta'=4$ and $\theta=\pi/2$, where the Fradkin-Shenker model belongs to the non-trivial SPT phase, the ES has an extensive $2^N$-fold degeneracy. However, the degeneracy splits immediately, as soon as we move away from $\theta'=\pi/2$, i.e., when the exact 1-form Wilson loops symmetry is explicitly broken.

We thus find that while the Higgs phase with the exact 1-form Wilson loop symmetry has an extensively degenerate ES, which is lifted by an arbitrarily small perturbation that explicit break the exact 1-form Wilson loop symmetry, and no transition in the ES can be observed at $\theta'\neq\pi/2$. While this is natural from the usual SPT perspective~\cite{Ruben_Gauging_kitaev_2021,quotient_SPT_2021}, it is somewhat at odds with the expectations based on the study of the physical rough boundary in Ref.~\cite{Ruben_Higgs_SPT}. There, a 2-fold degeneracy was found in the Higgs phase, which survives for perturbations that explicit break the exact 1-form Wilson loop symmetry and disappears at a physical boundary critical point. Thus, the ES does not mimics the physics of the specific boundary condition investigated in Ref.~\cite{Ruben_Higgs_SPT}. 
As we shall see, the extensive ES degeneracy is a direct consequence of the combination of the exact Gauss law constraint and the exact 1-form Wilson loop symmetry, thus it is impossible to reproduce the physical boundary transition of Ref.~\cite{Ruben_Higgs_SPT} within the ES of the Fradkin-Shenker model in the presence of the exact Gauss law constraint.

\subsubsection{Breaking exact Gauss law constraint}\label{ES_TC_cluster}

To further investigate the discrepancy between our ES degeneracy and the boundary degeneracy observed in Ref.~\cite{Ruben_Higgs_SPT}, we will study the ES of the emergent $\mathbb{Z}_2$ lattice gauge theory in the non-trivial SPT phase and the vicinity regime, where the exact Gauss law constraint is broken by a finite coupling $h_z$. As we will see, this not only has the effect of removing the extensive part of the ES degeneracy but, in the process, it also makes the remaining 2-fold ES degeneracy from the non-trivial SPT order more robust against perturbations which explicitly break the exact 1-form Wilson loop symmetry.

\begin{figure}
    \centering
    \includegraphics[scale=0.5]{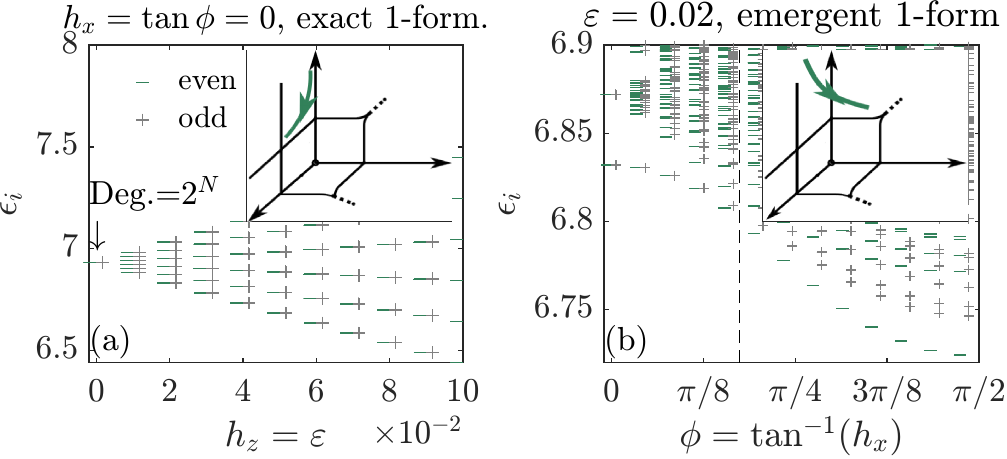}
    \caption{\textbf{ES near the fixed point of non-trivial SPT state.}  ES $\{\epsilon_i\}$ ($\epsilon_0\ge\epsilon_1\ge \epsilon_2\ge\cdots$) calculated by expressing the perturbed cluster state in Eq.~\eqref{perturbed_cluster} in terms of a PEPS, where the system is on an infinitely long cylinder with a finite circumference $N=10$, the PEPS transfer matrix fixed point is calculated using exact diagonalization. The paths are parameterized via $h_z=\varepsilon\cos(\phi), h_x=\tan\phi$ and $J=\varepsilon^{-1}\cos^{-1}\phi$. Insets show schematic paths of each panel in the phase diagram displayed in Fig.~\ref{Entanglement_cut}e. The ES levels are labeled by parity of the operator $\prod_iX_i$.
    (a) ES along a path satisfying $h_x=\tan\phi=0$, where the cluster model in a field has the exact 1-form Wilson loop symmetry. The even and odd parity levels are degenerate. 
    (b) ES  along a path satisfying $\delta=0.02$, When $\phi>0$, the 1-form Wilson loop is emergent, and we find the lowest even and odd parity levels of the ES to be degenerate (there is a exponentially small splitting with $N$) up to $\phi<\tan^{-1}(1/2)$, determined by the perturbative EH in Eq.~\eqref{EH_SPT} and denoted by the dashed line.} 
    \label{ES_cluster}
\end{figure}

We begin by considering what happens when we move slightly away from the fixed point of the non-trivial SPT phase. We consider a parametrization of a line in the three-dimensional parameter space $(h_x,h_z,J)$ using two parameters $\varepsilon$ and $\phi$: $h_z=\varepsilon\cos(\phi), h_x=\tan\phi$ and $J=\varepsilon^{-1}\cos^{-1}\phi$. A schematic of such a path can be found in the insets of Fig.~\ref{ES_cluster}a for $\phi=0$ and Fig.~\ref{ES_cluster}b for $\phi\neq 0$. When $\varepsilon$ is small, we obtain the ground state from perturbation theory:
\begin{align}\label{perturbed_cluster}
    \ket{\Psi_{\tcl}}=&\Biggl\{1+\frac{\varepsilon}{2}\sum_e \left[\frac{\cos (\phi) }{2}Z_e+\sin(\phi) X_e\right]+\frac{\varepsilon\cos\phi}{8}\sum_vX_v\notag\\
    &+O(\varepsilon^2)\Biggr\}\ket{\tcl},
\end{align}
where $\ket{\tcl}$ is the fixed point ground state of the non-trivial SPT phase shown in Eq.~\eqref{cluster_state}.

From $\ket{\Psi_{\tcl}}$, the EH can be derived: $H_{E,{\tcl}}=-\log(\rho_{{\tcl}})$, where the reduced density matrix is $\rho_{\tcl}=\Tr_L\ket{\Psi_{\tcl}}\bra{\Psi_{\tcl}}$. The degrees of freedom defining the EH are qubits in the right part $R$. For gapped phases, the qubits far from the entanglement cut are typically not relevant to entanglement, and only those near the entanglement cut are relevant. We can make this point precise by explicitly constructing a transformation that brings the original EH to the effective EH, which we now do.    

\begin{figure}
    \centering
    \includegraphics{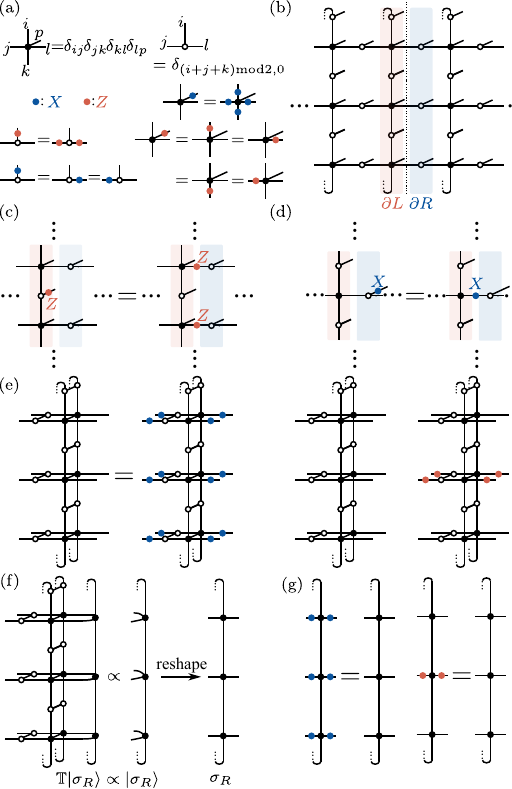}
    \caption{\textbf{Deriving the EH by perturbing the PEPS of the cluster state.}  (a) The definition of the PEPS tensors and their symmetries. (b) The PEPS represents the cluster state in Eq.~\eqref{cluster_state} on an infinitely long cylinder, and the effective EH acts on degrees of freedom on virtual bonds along the entanglement cut. (c) and (d) The terms in the effective EH in Eq.~\eqref{EH_SPT} can be derived by promoting $Z$ in $\partial L$ and $X$ in $\partial R$ from the physical level to the virtual level using the symmetry of the tensors in (a). (e) Transfer matrix of the PEPS and its symmetries. (f) The eigen-equation for the right dominant eigenvector $\ket{\sigma_R}$ of the transfer matrix. The right dominant eigenvector $\ket{\sigma_R}$ can be reshaped into a matrix $\sigma_R$. It is similar for the left dominant eigenvector $\sigma_L$. (g) The transfer matrix fixed point inherits the symmetry of the transfer matrix.}
    \label{Figure_Higgs_TN}
\end{figure}

The fixed point ground state $\ket{\tcl}$ shown in Eq.~\eqref{cluster_state} can be exactly expressed as a PEPS with a bond dimension two, as shown in Figs.~\ref{Figure_Higgs_TN}a and b. Then, we can use an isometry $\mathcal{V}_{\tcl}$ which is nothing but the right half part of the PEPS (see details in Appendix.~\ref{Appendix_EH}) to transform the $H_{E,\tcl}$ into an effective low energy EH $H^{\text{eff}}_{E,\tcl}$ that lives in the virtual degrees of freedom of the PEPS:
\begin{equation}
H^{\text{eff}}_{E,\tcl}=\mathcal{V}_{\tcl}H_{E,\tcl}\mathcal{V}^{\dagger}_{\tcl}+O(\varepsilon^2),
\end{equation}
which correctly captures the low-energy eigenvalues of the reduced density matrix at the order $O(\varepsilon)$~\cite{Cirac_2011}. In Appendix.~\ref{Appendix_EH}, we show that only the terms $\sum_{ e\in \partial R}X_e$ and $\sum_{e\in \partial L}Z_e$  in $\ket{\Psi_{\tcl}}$ nearest to the entanglement cut contribute the leading terms of $H^{\text{eff}}_{E,\tcl}$, where $\partial L$ ($\partial R$) is a set of edges at the boundary of region $L$ ($R$), see Fig.~\ref{Entanglement_cut}a and Fig.~\ref{Figure_Higgs_TN}b. The term $\sum_{v}X_v$, on the other hand, has no contribution to the leading terms of $H^{\text{eff}}_{E,\tcl}$. As shown in Figs.~\ref{Figure_Higgs_TN}c and d, by promoting $\sum_{ e\in \partial R}X_e$ and $\sum_{e\in \partial L}Z_e$ from the physical bonds to the virtual bonds along the entanglement cut, these terms become $\sum_iX_iX_{i+1}$  and $\sum_iZ_i$, respectively, in $H^{\text{eff}}_{E,\tcl}$.
Therefore, we arrive at the following effective EH from the perturbed cluster state in Eq.~\eqref{perturbed_cluster}:
\begin{equation}\label{EH_SPT}
H^{\text{eff}}_{E.\tcl}=-\frac{\varepsilon\cos\phi}{2}\sum_{i}Z_iZ_{i+1}-\varepsilon\sin\phi\sum_iX_i+N\log(2)+O(\varepsilon^2).
\end{equation}

Thus the EH takes the form of a 1-dimensional transverse field Ising model. At $\phi=0$ ($h_x=0$), when the emergent $\mathbb{Z}_2$ lattice gauge theory has the exact Wilson loop symmetry, the EH is in its ferromagnetic fixed-point with an exact $2$-fold degeneracy in the entire ES. At sufficiently small values of $\phi$, the EH remains in the ferromagnetic phase, and consequently the lowest ES level degeneracy survives, up to corrections that decay exponentially with the length $N$ of the entanglement cut. This lowest ES level degeneracy is only lifted when the EH undergoes an Ising phase transition at some finite value of $\phi$. This is similar to the boundary phase transition observed in Ref.~\cite{Ruben_Higgs_SPT}; however, to see it in the ES, we had to remove the exact Gauss law constraint.

The conclusions of this perturbative argument are also confirmed numerically in Figs~\ref{ES_cluster}a and b. The data shown is calculated by expressing the perturbed cluster state in Eq.~\eqref{perturbed_cluster} in terms of a PEPS, see details in Appendix~\ref{Appendix_iPEPS}. We find that when the $h_z$ term, which violates the exact Gauss law constraint, is turned on, the unnecessary $2^{N-1}$ fold degeneracy is lifted from the ES, but a $2$-fold exact degeneracy of the ground state remains (see Fig.~\ref{ES_cluster}a). Moreover, we find that this $2$-fold ES degeneracy is indeed robust to small perturbations that break the exact 1-form Wilson loop symmetry, see Fig.~\ref{ES_cluster}b, in contrast with the ES of the Fradkin-Shenker model with the exact Gauss law constraint in Fig.~\ref{fig:ES_Higgs_conf}b. We summarize the degeneracy in the ES of in the non-trivial SPT phase and Higgs regime as well as the volume between them  in Table.~\ref{deg_ES_SPT}.

From the perspective of SPT phases, the phenomenon observed here is quite unusual: the entanglement features associated with the non-trivial SPT phase survive even when the protecting exact 1-form symmetry is explicitly but weakly broken. We attribute this to the fact that the symmetry in question is a higher-form symmetry, which can exist as an emergent symmetry even when it is broken at the microscopic level~\cite{iqbal2022mean,Wen_emergent_high_form_2023,serna2024worldsheet}, and can thus retain the degeneracy in the low energy ES. However, we note that making precise sense of this idea is not obvious and requires further investigation. When tuning far away from $h_x=0$, we find a lifting of the two-fold degeneracy in the lowest ES level through a thermodynamics phase transition. We also note that a similar phenomenon can occur even in 0-form SPT phases, where it was observed in the context of ``quotient'' SPT phases~\cite{Fragility_SPT_2015,quotient_SPT_2021}.

\begin{table}[htbp]
    \centering
    \caption{Summary of the ES degeneracy in the non-trivial SPT phase, Higgs regime as well as the volume between them.}
    \label{deg_ES_SPT}
    \begin{tabular}{ccc}
    \toprule
    ES deg. & Exact 1-form Wilson  & Emergent 1-form Wilson \\
    \midrule
    Exact Gauss law & $2^N$  & 1 \\
    Emergent Gauss law & 2 & 2 or 1 \\
    \bottomrule
    \end{tabular}
\end{table}

\subsection{Understanding the degeneracies}

In this section, we show that the main features of the ES we found above follow from more generic symmetry considerations. First, we prove that the combination of the exact Gauss law constraint with the exact 1-form Wilson loop symmetry necessarily leads to an extensive, $2^{N-1}$-fold ES degeneracy independent of any SPT properties. Moreover, we show that a $2$-fold ES degeneracy is a characteristic feature of the non-trivial SPT order with an exact 1-form Wilson loop symmetry and a 0-form Ising symmetry (similarly to the known 2-fold ES degeneracy of the 1-dimensional non-trivial $\mathbb{Z}_2 \times \mathbb{Z}_2$ SPT phase~\cite{Pollmann_2011}). In the Higgs phase of the Fradkin-Shenker model with the exact 1-form Wilson loop symmetry, the $2^{N-1}$-fold non-generic ES degeneracy and the $2$-fold ES degeneracy associated with the non-trivial SPT order combine to produce an ES with $2^N$ degeneracy, while in a generic point in the non-trivial SPT phase (but with exact Gauss law constraint removed), only the $2$-fold ES degeneracy remains, making it robust to terms explicitly breaking the exact 1-form Wilson loop symmetry. We will now discuss these findings in detail.

\subsubsection{Origin of the extensive $2^{N-1}$-fold ES degeneracy}

To prove the extensive ES degeneracy in the presence of the exact Gauss law constraint and the exact 1-form Wilson loop symmetry, we first show that, as a consequence of exact Gauss law, $\rho_{\tGH}$ has a block-diagonal structure with $2^{N}$ sub-blocks. We will then show that $2^{N-1}$ sub-blocks have the same spectrum when there is the exact 1-form Wilson loop symmetry.

\begin{figure}
    \centering
    \includegraphics{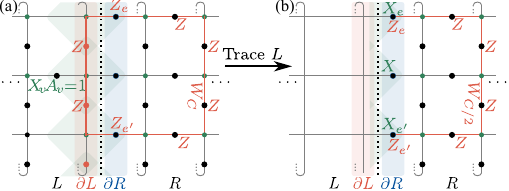}
    \caption{\textbf{Extensive ES degeneracy from the exact 1-form Wilson loop symmetry and the exact Gauss law constraint.} (a) A state with the exact Gauss law constraint and the exact 1-form Wilson loop symmetry (b) By tracing the left part $L$, the reduced density matrix has the local symmetry generated by the single $X$ operators in $\partial R$ and the open Wilson loop operators, which we denote as $W_{C/2}$.} 
    \label{fig:extensive_deg}
\end{figure}

Because of the exact Gauss law constraint 
any reduced density matrix of the Fradkin-Shenker model satisfies $[\rho_{\tGH}, X_e]=0$,  $\forall e\in \partial R$, see Fig.~\ref{fig:extensive_deg}, and is block diagonal with $2^N$ subblocks:
\begin{equation}\label{RDM_block_diagonal}
\rho_{\tGH}=\bigoplus_{\pmb{x}}p_{\pmb{x}}\rho_{\tGH,\pmb{x}},\quad p_{\pmb{x}}=\Tr \left(P_{\pmb{x}}\rho_{\tGH}\right),\quad\rho_{\tGH,\pmb{x}}=\frac{\rho_{\tGH} P_{\pmb{x}}}{p_{\pmb{x}}},
\end{equation}
where $P_{\pmb{x}}=2^{-N}\prod_{e\in \partial R}(1+x_e X_e)$ (with $x_e=\pm1$) is a projector to a specific sub-block labelled by $\pmb{x}=\{x_e|e\in \partial R\}$.

When the ground state has the exact 1-form Wilson loop symmetry (Fig.~\ref{fig:extensive_deg}a), $W_{C}\ket{\Psi_{\tGH}}=\ket{\Psi_{\tGH}}, \forall C$, the reduced density matrix has the symmetry of open Wilson loops denoted as $W_{C/2}$: $[W_{C/2},\rho_{\tGH}]=0, \forall C/2$, where $C/2$ is an open loop with two endpoints $e,e'\in\partial R$ along the entanglement cut, see Fig.~\ref{fig:extensive_deg}b. Because the open Wilson loop  symmetry $W_{C/2}$ relates different sub-blocks of $\rho_{\tGH}$, we have
\begin{equation}\label{sub_block_transformation}
    W_{C/2}P_{\pmb{x}}\rho_{\tGH} W^{\dagger}_{C/2}=P_{\pmb{x}'}W_{C/2}\rho_{\tGH} W^{\dagger}_{C/2}=P_{\pmb{x}'}\rho_{\tGH},
\end{equation}
where $|\pmb{x}'\rangle=\prod_{e\in(\partial R\cap C)}Z_{e}\ket{\pmb{x}}$, the subblocks $\pmb{x}$ and $\pmb{x'}$ are related by a unitary transformation: $\rho_{\tGH,\pmb{x}}=W_{C/2}\rho_{\tGH,\pmb{x}'}W_{C/2}$ ($p_{\pmb{x}}=p_{\pmb{x}'}$), and they therefore have the same spectrum. Since $[W_{C/2},\prod_{e\in \partial R} X_e]=0$, $\ket{\pmb{x}}$ and $\ket{\pmb{x}'}$ have the same parity, i.e., $\prod_{e\in\partial R}x_e=\prod_{e\in\partial R}x_e'$. In other words, each open Wilson loop has two end points on $\partial R$, which leaves the parity of $\ket{\pmb{x}}$ invariant. Therefore the spectra of the blocks with the same parity are identical, and the ES of $\rho_{\tGH}$ is $2^{N-1}$-fold degenerate.

\subsubsection{Origin of the 2-fold ES degeneracy in the non-trivial SPT phase}

We now show that the 2-fold degeneracy of the ES is a universal property of the non-trivial SPT phase with exact 1-form Wilson loop symmetry and 0-form Ising symmetry. For 1-dimensional non-trivial SPT phases, the necessary ES degeneracy of ground states can be derived using the symmetries of the tensors of the matrix product states (MPS)~\cite{Pollmann_2011}. We can similarly derive the ES degeneracy in the non-trivial SPT phase in the $h_x=0$ plane from the symmetries of the tensors in a PEPS representation of the ground states. From the PEPS of the cluster state, we find that the tensor has the symmetry shown in Fig.~\ref{Figure_Higgs_TN}a, from which the 0-form Ising symmetry and the 1-form exact Wilson loop symmetry can be derived. It is reasonable to assume that any PEPS tensor in the non-trivial SPT phase with the exact 1-form Wilson loop symmetry has the same symmetries as the fixed point PEPS tensor (although the representations can be different). This is because we expect that any symmetric deformation or perturbation does not change the symmetries of the PEPS tensors as long as no phase transition occurs, which is a principle used to classify gapped phases using tensor network states~\cite{Xie_chen_2011_1,Xie_chen_2011_2,Schuch_2011_classify}.

To make this precise, we first analyze the symmetry of the transfer matrix $\mathbb{T}$ of the PEPS, as shown in Fig.~\ref{Figure_Higgs_TN}e. Using the relations from Fig.~\ref{Figure_Higgs_TN}a, we can read off that
\begin{align}
    &\left(\prod_i X_i\otimes\prod_i X_i\right) \mathbb{T} \left(\prod_i X_i\otimes\prod_i X_i\right)= \mathbb{T}\notag \\
     &\left(Z_j\otimes Z_j\right) \mathbb{T} \left(Z_j\otimes Z_j\right)=\mathbb{T},\quad \forall j.
\end{align}
As the system is in an SPT phase, the left and right dominant eigenvector $\bra{\sigma_L}$ and $\ket{\sigma_R}$ of the transfer matrix $\mathbb{T}$ (see Fig.~\ref{Figure_Higgs_TN}f) are unique. As a consequence also the left and right eigenvectors of the transfer matrix inherit these symmetries: $(\prod_i X_i \otimes \prod_i X_i)\ket{\sigma_R}=\ket{\sigma_R}$ and $ Z_j \otimes Z_j\ket{\sigma_R}=\ket{\sigma_R}, \forall j$, as shown in  Fig.~\ref{Figure_Higgs_TN}g, and the similarly for $\bra{\sigma_L}$. For notational convenience, we reshape these eigenvectors to matrices again leading to the symmetries:
\begin{equation}\label{eq:Higgs_ES_deg}
\left(\prod_i X_i\right) \sigma_{L(R)} \left(\prod_i X_i\right) = \sigma_{L(R)}, \quad Z_j \sigma_{L(R)} Z_j = \sigma_{L(R)},\quad  \forall j.
\end{equation}
The reduced density matrix is then obtained from $\rho\sim\sqrt{\sigma^{\text{T}}_L}\sigma_R\sqrt{\sigma^{\text{T}}_L}\sim \sigma^{\text{T}}_L\sigma_R$~\cite{Cirac_2011} (``$\sim$'' means two operators have the same non-zero eigenvalues). Following the individual symmetries, $\sigma_L\sigma_R^{\text{T}}$ has a global symmetry $\prod_iX_i$ and a local symmetry $Z_j$ acting on the entanglement degrees of freedom in 1 dimension, which anti-commute with each other. Therefore, every eigenvalue of $\rho\sim \sigma_L\sigma_R^{\text{T}}$, has to be at least two-fold degenerate. We conclude that the 2-fold necessary degeneracy in the ES is a universal property of the non-trivial SPT phase with the exact 1-form Wilson loop and the 0-form Ising symmetry.

We thus find that there are two separate sources of the extensive ES degeneracy in the Fradkin-Shenker model with the exact 1-form Wilson loop symmetry. First, from the proof in the previous section, the $2^{N-1}$-fold ES degeneracy arises from the interplay between the exact Gauss law constraint and the exact 1-form Wilson loop symmetry. Second, in the Higgs phase, there is an additional 2-fold ES degeneracy associated with the non-trivial SPT order, resulting in the observed total $2^N$-fold ES degeneracy.

\subsubsection{Robustness of the 2-fold ES degeneracy against explicit 1-form symmetry breaking perturbations}

Finally, let us return to the issue of the robustness of the $2$-fold ES degeneracy when explicitly breaking the exct 1-form Wilson loop symmetry. The conditions in Eq.~\eqref{eq:Higgs_ES_deg} suggest that the effective EH has a global symmetry $\prod_iX_i$ and a local symmetry $Z_i,\forall i$. In the absence of additional constraints or fine-tuning, a generic EH with these properties will be somewhere deep in a phase spontaneously breaking $\prod_iX_i$ due to the anti-commutation relation between the two symmetries.

A bulk perturbation that explicitly breaks the exact 1-form Wilson loop symmetry but preserves the 0-form Ising symmetry then modifies the EH by additional $\prod_iX_i$ symmetric local terms. 
As a result, the EH will no longer have the local symmetry generated by $Z_i$, removing the exact degeneracies. However, for sufficiently weak perturbations, the EH still remains in the phase where the symmetry $\prod_iX_i$ is spontaneously broken. We thus expect that when weakly breaking the exact 1-form Wilson loop symmetry of a generic point within the non-trivial SPT phase, the lowest level of ES continues to have a 2-fold degeneracy, which is no longer exact, but is exponentially small in the length of the entanglement cut. This expectation was already observed in the EH in Eq.~\eqref{EH_SPT} and in our numerics shown in Fig.~\ref{ES_cluster}b.

In contrast, if we consider a bulk perturbation breaking the 0-form spin flip symmetry, e.g., $\sum_v Z_v$, it modifies the effective EH by additional local terms such as $\sum_iZ_i$, which explicitly break the symmetry $\prod_iX_i$. Thus, because a usual spontaneous symmetry-breaking phase is not robust against explicit symmetry-breaking perturbations, weakly breaking the bulk 0-form Ising symmetry will lift the 2-fold degeneracy of the lowest ES level.

While the argument for robustness applies to generic points in the SPT phase, it does not rule out fine-tuned points where the degeneracy becomes unstable to arbitrarily weak perturbations. This is exactly what happens in the presence of the exact Gauss law constraint, which precludes the terms in the effective EH that spontaneously break the symmetry $\prod_iX_i$. Instead, in this case, we end up with an effective EH which is almost equal to an identity matrix with no stability against perturbations.

\section{Entanglement Hamiltonian and entanglement spectrum in the topologically ordered phase}\label{Sec:EH_ES_TC}
\begin{figure}
    \centering
    \includegraphics{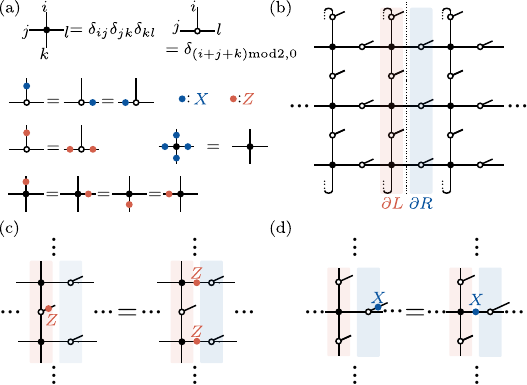}
    \caption{\textbf{Deriving EH by perturbing the PEPS of the fixed point TC model.}  (a) The definition of the PEPS tensors and their symmetries. (b) The PEPS on an infinitely long cylinder and the entanglement cut. (c) and (d) The terms in the EH in Eq.~\eqref{ent_ham_all} are derived by promoting the $Z$ ($X$) in $\partial L$ ($\partial R$) from the physical level to the virtual level using the symmetry of the tensors in (a).}
    \label{Fig_EH_TC}
\end{figure}
In this section, we consider the ES of the topologically ordered phase, where the 1-form symmetries break spontaneously, and contrast it with the ES near the non-trivial SPT fixed point discussed in the previous section. We find that the EHs have similar local terms in the two cases. However, because of the spontaneous 1-form symmetry breaking, the EH of a topologically ordered state contains a non-local projector~\cite{Cirac_2011}, generally resulting a non-degenerate ES.

We again consider a cylinder geometry and take a bipartition of the system into two half-cylinders, separated by an entanglement cut far from either boundary, which is convenient for numerical calculation. Since this entanglement cut is a non-contractible loop, the reduced density matrix depends on the choice of the topologically degenerate ground state. In contrast, if the entanglement cut is a contractible loop, the reduced density matrix is independent of the choice of the topologically degenerate ground states. It is known that when the ground state is chosen as the trivial minimally entangled states (MES)~\cite{MES_2012}, i.e., a simultaneous eigenstate of the Wilson loop $W_{C_y}$ and 't Hooft loop $T_{\hat{C}_y}$ shown in and Fig.~\ref{Entanglement_cut}d, with eigenvalues $(1,1)$, the reduced density matrix is qualitatively the same as the one obtained from a contractible entanglement cut. We always consider the entanglement priorities in the $\mathbb{Z}_2$ topologically ordered phase using the trivial MES to mimic a contractible entanglement cut. Moreover, Sec.~\ref{sec:EH_ES_SPT},
we will again use the perturbed PEPS to calculate ES in the topologically ordered phase.

\begin{figure}
    \centering
    \includegraphics[scale=0.5]{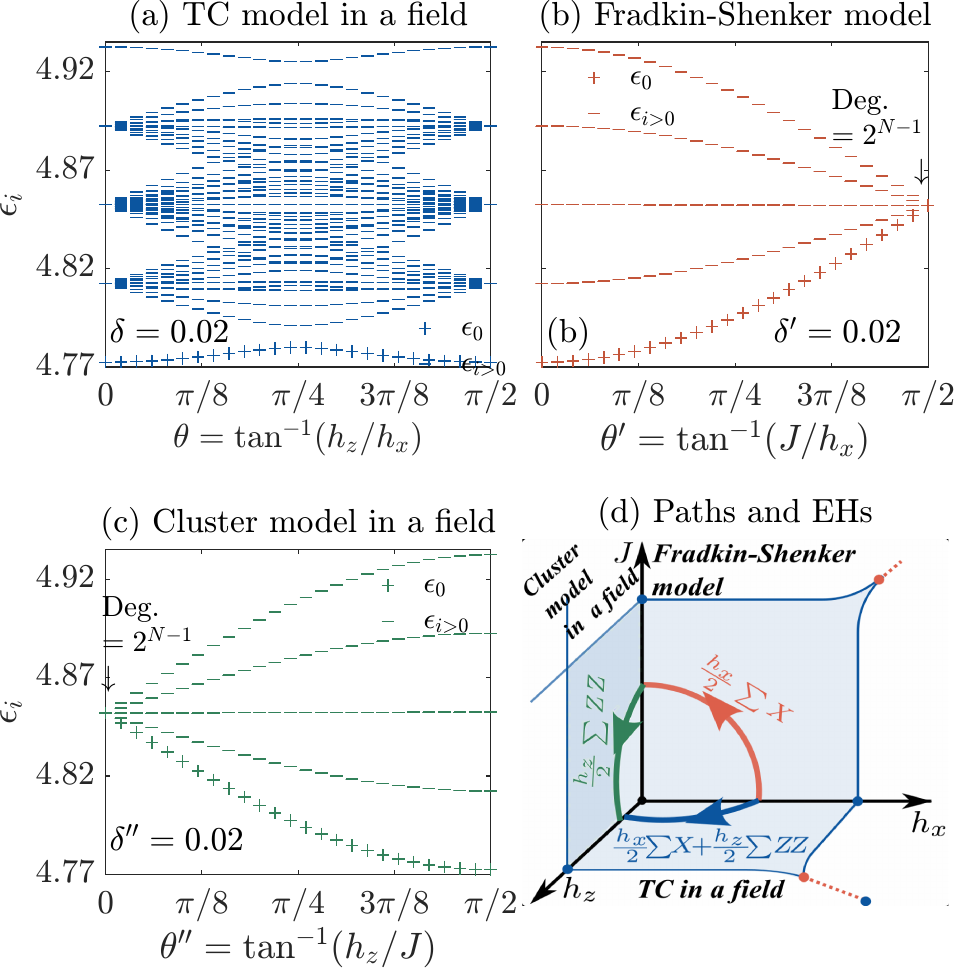}
    \caption{\textbf{ES in the topologically ordered phase.} ES $\{\epsilon_i\}$ ($\epsilon_0\ge\epsilon_1\ge \epsilon_2\ge\cdots$) of the trivial MES on an infinitely long cylinder with a finite circumference $N=8$, 
    which is calculated by expressing the perturbed TC state in Eq.~\eqref{eq:perturbed_MES} in terms of a PEPS, and the PEPS transfer matrix fixed point is calculated using exact diagonalization.
    (a) ES of the TC model in a field along $\delta=\sqrt{h_x^2+h_z^2}=0.02$. (b) ES of the Fradkin-Shenker model along $\delta'=\sqrt{h_x^2+J^2}=0.02$.  (c) ES of the cluster model in a field along $\delta''=\sqrt{h_z^2+J^2}=0.02$. (d) Schematic paths in the phase diagram displayed 
    along which the ES in the panels (a), (b), and (c) are calculated.  The leading local terms of the EHs are also shown.}
    \label{ES_toric_code}
\end{figure}

At the fixed point of the topologically ordered phase ($h_x=h_z=J=0$), one ground state can be written as (ignoring the product state $\prod_v \ket{+}_v$ on vertices):
\begin{equation}\label{fixed_point_TC}
    \ket{\tTC}=\prod_{v}\frac{1+A_v}{\sqrt{2}}\prod_e\ket{0}_e.
\end{equation}
It has an exact representation in terms of a PEPS~\cite{single_double_line_2008}, as shown in Figs.~\ref{Fig_EH_TC}a and b. The PEPS of the cluster state $\ket{\tcl}$ defined in Figs.~\ref{Figure_Higgs_TN}a and b is very similar to that of the $\ket{\tTC}$, but former has extra physical degrees of freedom at each vertex, corresponding to the matter fields in the gauge theory. The two are related via $\ket{\tTC}=2^{\#v/2}\prod_v\bra{+}_v\ket{\tcl}$, where $\#v$ is the total number of vertices. The PEPS $\ket{\tTC}$ is invariant under the non-contractible Wilson loop symmetry $W_{C_y}\ket{\tTC}=\ket{\tTC}$ but not invariant under the non-contractible 't Hooft loop symmetry $T_{\hat{C_y}}\ket{\tTC}\neq\ket{\tTC}$. The trivial MES can be obtained by applying a projector: $(1+T_{\hat{C}_y})\ket{\tTC}/\sqrt{2}$.

We now consider small perturbations near the fixed point $(h_x,h_z,J)=(0,0,0)$. We obtain the corresponding trivial MES
\begin{align}\label{eq:perturbed_MES}
  \ket{\Psi_{\ttop}}&=\left[
  1+\frac{h_x}{4}\sum_e X_e+\frac{h_z}{4}\sum_e Z_e+\frac{J}{4}\sum_{\langle vev'\rangle} Z_vZ_eZ_{v'}+O(\delta_{r}^2)
  \right]\notag\\
  &\times\frac{1 +T_{\hat{C}_y}}{\sqrt{2}}\ket{\tTC}\otimes \prod_v\ket{+}_v,
\end{align}
where $\delta_{r}=\sqrt{h_x^2+h_z^2+J^2}$ is small. From $\ket{\Psi_{\ttop}}$, we can obtain the EH: $H_{E,{\ttop}}=-\log(\rho_{{\ttop}})$. Similar to the derivation of the EH of the perturbed cluster state in Sec.~\ref{sec:EH_ES_SPT}, we perform a transformation to derive the effective EH and find that only the terms $\sum_{ e\in \partial R}X_e$ and $\sum_{e\in \partial L}Z_e$  in $\ket{\Psi_{\ttop}}$ nearest to the entanglement cut contribute to the leading terms of the effective EH $H^{\text{eff}}_{E,\ttop}$. Here, $\partial L$ ($\partial R$) is a set of edges at the boundary of the region $L$ ($R$), see Fig.~\ref{Fig_EH_TC}b. As shown in Figs.~\ref{Fig_EH_TC}c and d, by promoting $\sum_{ e\in \partial R}X_e$ and $\sum_{e\in \partial L}Z_e$ from the physical bonds to the virtual bonds along the entanglement cut, these terms become $\sum_iZ_iZ_{i+1}$  and $\sum_iX_i$ in $H^{\text{eff}}_{E,\ttop}$. Another similarity is that the gauge-matter coupling term $\sum_{\langle vev'\rangle}Z_vZ_eZ_{v'}$ has no contribution to the leading terms of $H^{\text{eff}}_{E,\ttop}$ (see details in Appendix.~\ref{Appendix_EH}).
However, the important difference between the two cases arises from the projector $(1+T_{\hat{C}_y})/\sqrt{2}$ onto the even sector of the 't Hooft 1-form symmetry, resulting from the trivial MES. This projector becomes $P_{+}\equiv(1+\prod_{i}X_{i})/2$ on the virtual bonds of the PEPS. Taking it into consideration,
we find that the effective EH of the trivial MES reads
\begin{equation}\label{ent_ham_all}
H^{\text{eff}}_{E,\ttop}=P_{+}\left[\log2^{N-1}-\sum_{i=1}^{N}\left(\frac{h_z}{2}Z_iZ_{i+1}+\frac{h_x}{2}X_i\right)+O(\delta_r^2)\right].
\end{equation}
Crucially, the gauge-matter coupling $J$ does not appear in the leading order of $H^{\text{eff}}_{E,\ttop}$. Therefore, we obtain for the Fradkin-Shenker model ($h_z = 0$) and the cluster model in a field ($h_x = 0$), the following leading terms of the EH, which are all commuting:
\begin{align}
    H^{\text{eff}}_{E,\tGH}&=P_{+}\left[\log2^{N-1}-\sum_{i=1}^{N}\frac{h_x}{2}X_i+O(\delta^{\prime 2})\right],\label{EH:IGH_in_topo}\\
    H^{\text{eff}}_{E,\tcl}&=P_{+}\left[\log2^{N-1}-\sum_{i=1}^{N}\frac{h_z}{2}Z_iZ_{i+1}+O(\delta^{\prime\prime 2})\right],
\end{align}
where $\delta'=\sqrt{h_x^2+J^2}$ and $\delta''=\sqrt{h_z^2+J^2}$.
We schematically illustrate the local terms in the EH on the three planes in the perturbative regime in Fig.~\ref{ES_toric_code}d.

The EH $H^{\text{eff}}_{E,\ttop}$ in Eq.~\eqref{ent_ham_all} is similar to that of the perturbed cluster model shown in Eq.~\eqref{EH_SPT}, which follows from the aforementioned relationship between the two fixed point ground states. However, the two cases differ by the fact that $H^{\text{eff}}_{E,\ttop}$ contains a projector $P_{+}$, stemming from the spontaneous 1-form symmetry breaking, which projects out ``half'' of ES corresponding to the energy spectrum of the 1-dimensional transverse field Ising model. Consequently, the ES is generically non-degenerate. However, when $h_x\approx h_z$ ($h_x=h_z,J=0$ in the TC model in a field because of the electric-magnetic duality), the EH is at an Ising phase transition point, and the ES is described by the Ising conformal field theory. This is consistent with a previous study of another self-dual $\mathbb{Z}_2$ topological state with a different entanglement cut~\cite{G_Vidal_2015}. 

We can complement the above perturbative analysis by computing the ES from the PEPS representing the perturbed ground state in Eq.~\eqref{eq:perturbed_MES}. The results for the trivial MES in the topologically ordered phase are shown in Figs.~\ref{ES_toric_code}a, b and c, which show the ES of the TC model in a field, the Fradkin-Shenker model and the cluster model in a field separately. We parameterize the phase diagram by three angles, $\theta, \theta', \theta''$, defined as $\tan{\theta} = h_z/h_x$, $\tan{\theta'} = J/h_x$ and $\tan{\theta''} = h_z/J$, respectively.
As expected, we find that, generically, the ES is non-degenerate.
Moreover, we observe that the ES of the TC model in a field in Fig.~\ref{ES_toric_code}a  shows a reduced gap at $\theta=\pi/4$. The remaining gap is due to finite size effect. However, we have also checked the momentum resolved ES, which is consistent with the prediction from the (1+1)-dimensional Ising conformal field theory. 

It is also worth comparing the ES of the TC model in a field (Fig.~\ref{ES_toric_code}a) with that of the Fradkin-Shenker model (Fig.~\ref{ES_toric_code}b). Unlike the ES of the TC model in a field, the ES of the Fradkin-Shenker model is not symmetric about $\pi/4$ (see also Fig.~\ref{fig:ES_Higgs_conf}b in comparison to Fig.~\ref{fig:ES_Higgs_conf}a), despite the bulk phase diagram of the Fradkin-Shenker model being symmetric about $\theta'=\pi/4$. The non-symmetric ES can be explained by combining the quantum channel that relates the two reduced density matrices with the electric-magnetic duality of the TC model in a field; see details in Appendix~\ref{app:why_ES_of_FS_non_symmetric}. 

We also note that the $2^N$-fold ES degeneracy we observed previously along the line $h_x=h_z=0$, where both the 1-form Wilson loop symmetry and the Gauss law constraint are exact, is reduced to a $2^{N-1}$-fold ES degeneracy (which as we derived is required by the symmetries) in the topologically ordered phase, see Figs.~\ref{ES_toric_code}b and c. This reinforces our previous analysis which explained the additional $2$-fold ES degeneracy is associated to the non-trivial SPT order, which is no longer present in the topologically ordered phase.

\section{Topological entanglement entropy and spontaneous 1-form symmetry breaking}\label{distillable_TEE_IGH}

One standard way to characterize topologically ordered phases is in terms of their topological entanglement entropy (TEE)~\cite{TEE_levin_Wen_2006,TEE_Kitaev_Preskill_2006}. Usually, this is defined as a characteristic constant contribution to the entropy of a sub-region. In this section, we first show that the spontaneous 1-form symmetry breaking of the topologically ordered phase is essential for the appearance of TEE. Then we take the Gauss law constraint, which motivates the study of distillable entanglement, into consideration, and explore whether the distillable EE of the Fradkin-Shenker model contains the TEE using the fact that the spontaneous 1-form symmetries breaking is essential for TEE.

\subsection{TEE and spontaneous 1-form symmetry breaking without Gauss law constraint}\label{TEE_and_1_form_SSB}

In this section, we show a direct connection between spontaneous 1-form symmetry breaking and TEE in a tensor product Hilbert space, neglecting the exact Gauss law constraint. As mentioned in the beginning of Sec.~\ref{Sec:EH_ES_TC}, we consider the trivial minimally entangled state (MES), such that our conclusion is general in the sense that it is valid for any contractible entanglement cut in any ground states. In principle, the TEE can be extracted from the $n$-R\'enyi entropy $S_n$ (where $n\rightarrow 1$ corresponds to the von Neumann entropy) as~\cite{MES_2012}: $S_n=\alpha_n N-\gamma$, where $N$ is the length of the entanglement cut, $\alpha_n$ is a non-universal constant and $\gamma$ is the universal TEE characterizing topological order~\cite{TEE_Kitaev_Preskill_2006,TEE_levin_Wen_2006,TEE_Renyi_wen_2009,MES_2012}. At the fixed point ($h_x=h_z=J=0$), it is easy to check that $S_n=N\log2-\log2$~\cite{EE_toric_code}, according to Eq.~\eqref{ent_ham_all}, so the TEE is $\gamma=\log 2$. However, away from the fixed point of the toric code, the ES is changed, but the TEE remains robust. How can one understand this behavior?

According to the EH in Eq.~\eqref{ent_ham_all}, the reduced density matrix of $\ket{\Psi_{\ttop}}$ should have the following form~\cite{Schuch_transfer_operator_2013}:
\begin{equation}\label{assumption}
\rho_{\ttop}\sim \frac{1+T_{\hat{C}_y}}{2}\exp(-H_E),
\end{equation}
where the low-energy space of $H_E$ is described by an effective 1-dimensional Hamiltonian along the entanglement cut consisting of a sum over local or quasi-local terms. $H_{E}$ satisfies $[T_{\hat{C}_y},H_{E}]=0$ but $T_{\hat{C}_y}H_{E}\neq H_{E}$ due to spontaneous t'Hooft loop symmetry breaking, i.e., $T_{\hat{C}_y}\ket{\tTC}$ and $\ket{\tTC}$ are orthogonal. 

Let us consider the $n$-R\'enyi entropy of a density matrix $\rho$:  $S_n(\rho)=(1-n)^{-1}\log\Tr\rho^n$.
The R\'enyi EE of $\rho_{\ttop}$ in Eq.~\eqref{assumption} is
\begin{align}\label{Renyi_EE}
S_n(\rho_{\ttop})&=\frac{1}{1-n}\log\left( \Tr \frac{e^{-nH_E}}{2}+ \Tr \frac{T_{\hat{C}_y}e^{-nH_E}}{2}\right)\notag\\
&-\frac{n}{1-n}\log\left( \Tr \frac{e^{-H_E}}{2}+ \Tr \frac{T_{\hat{C}_y}e^{-H_E}}{2}\right).
\end{align}
The second line accounts for the normalization.
Since $-\left(\log \Tr e^{-nH_E}\right)/n$ can be interpreted as the free energy of the quantum system $H_E$ at the temperature $1/n$, we have $-\log \Tr e^{-nH_E}=nNf(1/n)$ when the circumference $N$ is large enough, where $f$ is the free energy density. What about the other terms? $\Tr\left(T_{\hat{C}_y}e^{-nH_E}\right)/\Tr e^{-nH_E}$ is the expectation value of $T_{\hat{C}_y}$, which decays exponentially with the circumference $N$ and vanishes when $N\rightarrow\infty$, because of the spontaneous 1-form 't Hooft loop symmetry breaking.
Therefore, in Eq.~\eqref{Renyi_EE}, the terms $\Tr\left(T_{\hat{C}_y}e^{-nH_E}\right)$ and $\Tr\left(T_{\hat{C}_y}e^{-H_E}\right)$ are exponentially small compare to $\Tr\left(e^{-nH_E}\right)$ and $\Tr\left(e^{-H_E}\right)$, seperately, and they can be neglected. By substituting the free energy interpretation into Eq.~\eqref{Renyi_EE}, the R\'enyi EE can be simplified:
\begin{align}\label{REE_of_TC}
    S_n(\rho_{\ttop})&=-\frac{n N f(1/n)}{1-n}-\frac{\log(2)}{1-n}+\frac{n N f(1)}{1-n}+\frac{n\log2 }{1-n}  \notag\\ &=-\frac{1}{1/n-1}\left[f(1/n)-f(1)\right]N-\log2.
\end{align}
Here, we reproduced the result of Ref.~\cite{TEE_Renyi_wen_2009}, the sub-leading $-\log{2}$ contribution appearing in Eq.~\eqref{REE_of_TC} is exactly the expected TEE, but we here relate the TEE with spontaneous 1-form symmetry breaking. When $n\rightarrow 1$, it becomes the von Neumann EE $S_1(\rho_\ttop)=s(1)N-\log2$, where $s(1)$ is the von Neumann entropy density of the quantum system $H_E$ at the entanglement temperature $T_E=1$~\cite{Li_Haldane_2008}.  On the other hand, when we consider a ground state in which the 1-form symmetry is not broken spontaneously, which means that $T_{\hat{C}_y}H_E=H_E$ and  $\Tr\left(T_{\hat{C}_y}e^{-nH_E}\right)/\Tr e^{-nH_E}=1$, we conclude from Eq.~\eqref{Renyi_EE} that $S_n=-(1/n-1)^{-1}\left[f(1/n)-f(1)\right]N$, \emph{without} any correction to the area law. Therefore, if the 1-form 't Hooft loop symmetry does not break spontaneously, there will be no TEE, as expected. In summary, if $\rho_{\text{\ttop}}\sim (1+T_{\hat{C}_y})\exp(-H_E)/2\not\propto\exp(-H_E)$, where $H_E$ is a sum over local or quasi-local terms, then TEE is $\log2$.

\subsection{Distillable EE by imposing the exact Gauss law constraint}
In the Fradkin-Shenker model as well as other fundamental gauge theories, where the Gauss law constraint is exact, the definition of EE, and thus the TEE, becomes less obvious, as the physical Hilbert space lacks a tensor product structure. A possibility is to instead consider the \emph{distillable entanglement}, which is a measure of entanglement that can be adapted to take the exact Gauss law constraint into account. In this section, we investigate the behavior of the distillable EE in the deconfined phase of the  Fradkin-Shenker model, focusing on how the physics of the topological order (i.e., the SSB of 1-form symmetries) manifests itself in this quantity. We first discuss the definition of the distillable EE, and then show the numerical results for the Fradkin-Shenker model and analytical understanding of the numerical results using the derivations in Sec.~\ref{TEE_and_1_form_SSB}.
\subsubsection{Definition of the distillable EE}

In general, the distillable entanglement is defined in terms of the maximal number of Bell pairs $m$ that can be created, with high fidelity, from $n$ copies of a quantum state via LOCC operations; namely, the distillable entanglement is given by the ratio $m/n$ in the limit that $n\to\infty$. It can be shown that for pure states, this quantity equals the von Neumann EE~\cite{nielsen2010quantum}. The motivation for defining the distillable entanglement is to treat the quantum state as a resource, which can be turned into practically useful Bell pairs. From this perspective, it is natural, in some cases, to put additional physical restrictions on the operations one is allowed to execute in the process of preparing Bell pairs. In particular, if the system is described by a gauge theory, one should only be able to execute operations that obey gauge-invariance (satisfying the exact Gauss law constraint). One can then use the gauge invariant distillable entanglement as a \emph{definition} of entanglement, between some region $A$ and its complement, in the gauge theory.

\begin{figure}
    \centering
    \includegraphics{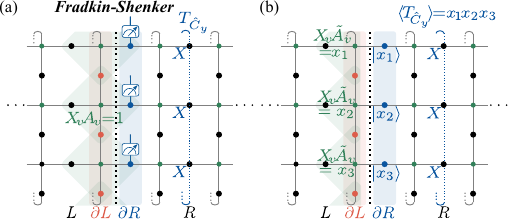}
    \caption{\textbf{Distilling entanglement from a quantum state constrained by the exact Gauss law.} (a) Measurement of the physical degrees of freedom in $\partial R$, and illustration of an  is an exact 1-form non-contractible 't Hooft loop operator $T_{\hat{C}_y}$. (b) The Gauss law constraint of the measured state does not cross the entanglement cut. The expectation value of the non-contractible exact 't Hooft 1-form symmetry operator is fixed by the measurement. As a consequence, the measured state can not spontaneously break the 1-form t' Hooft loop symmetry and the TEE can not be extracted from the measured state.
    }
    \label{figure_distillable}
\end{figure}

From an operational viewpoint, to extract the distillable EE, we can measure the gauge degrees of freedom along the entanglement cut using $X$ operators, see Fig.~\ref{figure_distillable}a. 
In the measured state, the degrees of freedom in $\partial R$ become a product state disentangled from the rest of the system, and the exact Gauss law is modified such that it does not cross the entanglement cut, as shown in Fig.~\ref{figure_distillable}b. The entanglement of the measured state can be calculated in the usual way. The reduced density matrix of the measured state $\ket{\Psi_{\pmb{x},\tGH}}=P_{\pmb{x}}\ket{\Psi_{\tGH}}/\sqrt{p_{\pmb{x}}}$ is just one of the subblocks $\rho_{\tGH,\pmb{x}}$ in Eq.~\eqref{RDM_block_diagonal}. 
Since the measurement results are random, one needs to measure sufficiently many times and calculate the usual EE of each measured state. It has been shown that the average of the von Neumann entanglement entropies $S_1$ of all measured states is the distillable EE~\cite{Hamma_2005,PRD_decompose_EE_2012,Distillable_entanglement_2016,Gauge_ent_1,Gauge_ent_2,Gauge_ent_3,Gauge_ent_4,Gauge_ent_5}:
\begin{equation}\label{eq:distillable_ee_measurements}
S_{D,1}=\sum_{\pmb{x}}p_{\pmb{x}}S_1(\rho_{\tGH,\pmb{x}}),
\end{equation}
which is just one part of the von Neumann EE $S_1$ of the full reduced density matrix $\rho_{\tGH}$~\cite{Wiseman_2003, PRD_decompose_EE_2012}:
\begin{equation}\label{distillable_von_Neumann_TEE}
   S_1(\rho_{\tGH})=-\sum_{\pmb{x}}p_{\pmb{x}}\log p_{\pmb{x}}+S_{D,1},
\end{equation}
where the first part is the classical Shannon entropy of the probability distribution $p_{\pmb{x}}=\bra{\Psi_{\tGH}}P_{\pmb{x}}\ket{\Psi_{\tGH}}$ of the measurement outcomes, and the second part is the distillable entanglement entropy $S_{D,1}$. 

From the first-order perturbed ground states of the Fradkin-Shenker model, one can show that the spectrum of each subblock $\rho_{\tGH,\pmb{x}}$ in Eq.~\eqref{RDM_block_diagonal} has a non-degenerate dominant eigenvalue and a large gap, and the low energy ES of the Fradkin-Shenker model in Figs.~\ref{fig:ES_Higgs_conf}b and \ref{ES_toric_code}b come from the largest eigenvalue of each subblock, so the low energy ES can be labeled by the subblock index $\pmb{x}$. Moreover, in Appendix.~\ref{Appendix_EH} we show that the subblock index labelled ES $\{\epsilon_{\pmb{x}}\}$ is related to the probability distribution via $p_{\pmb{x}}=\exp(-\epsilon_{\pmb{x}})+O(\delta'^2)$, where $\delta'^2=h_x^2+J^2$. Therefore, if we only consider the first-order classical EHs of Fradkin-Shenker model in Eqs.~\eqref{EH_GH_trivial} and ~\eqref{EH:IGH_in_topo}, we only obtain the Shannon entropy. The distillable EE can only be extracted from the higher-order perturbative terms.

Here we go beyond the leading order and compute the universal behavior of the distillable EE. In particular, we will next determine whether it contains the universal TEE contribution expected for topologically ordered phases.
To answer this question, we examine the presence or absence of the 1-form symmetry breaking in the measured state to determine whether the distillable EE contains the universal TEE. 
We will find that the TEE is \emph{absent} in the distillable EE when the 't Hooft loop symmetry is exact and appears only in the deconfined phase with emergent 1-form 't Hooft loop symmetry.

\subsubsection{Numerical results for the distillable EE}

\begin{figure}[t]
    \centering
    \includegraphics[scale=0.5]{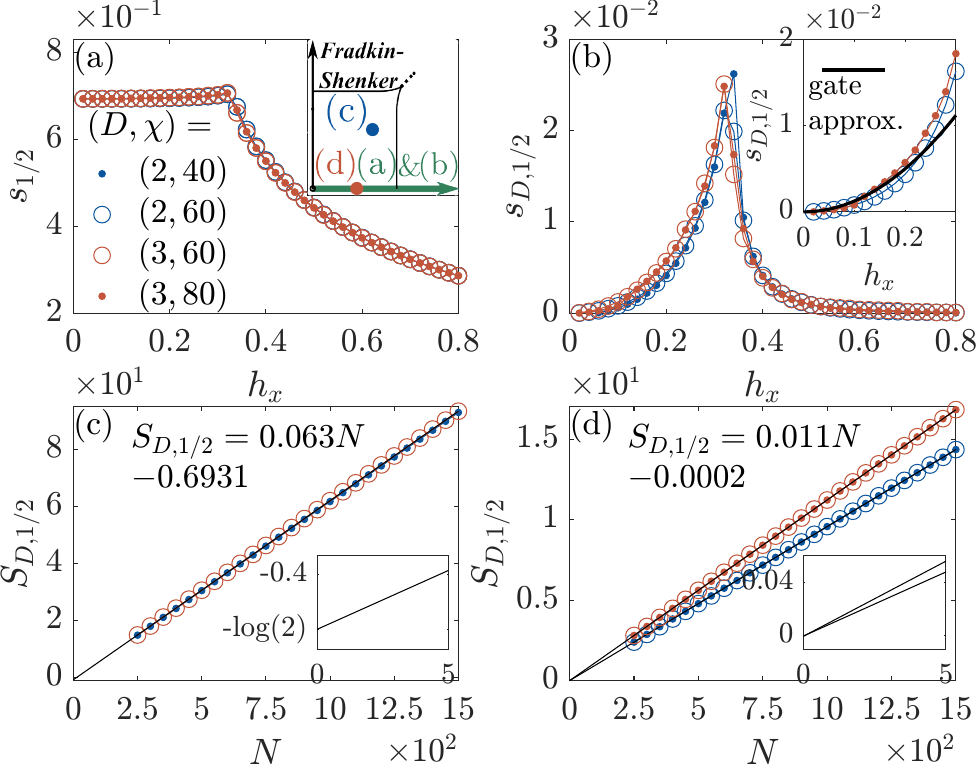}
    \caption{\textbf{Distillable Rényi EE of the Fradkin-Shenker model.} The results are extracted from the ground states approximated using the variationally optimized PEPS with a bond dimension $D$, and the boundary MPS of the PEPS has a bond dimension $\chi$. The legend in (a) shows $(D,\chi)$ and applies to all subplots. Inset in (a)
shows the schematic path or positions of each panel in the $(h_x,J)$ plane in the phase diagram. 
(a) Total 
    and (b) distillable 
    $1/2$-R\'enyi EE densities for a pure gauge theory ($J=h_z=0$) extrapolated to the thermodynamic limit $N\rightarrow\infty$. Inset of (b): Comparison with the results obtained from the gate decomposition approximation (shown in Sec.~\ref{sec:distillable_ent_non_zero}) in Eq.~\eqref{decompose_gate}.  (c)  Distillable R\'enyi EE  $S_{D,1/2}(\rho_\tGH)$ at $(h_x,h_z,J)\approx(0.0804,0,0.1226)$.
    Inset: zoom in of the extrapolation yielding the topological correction $\gamma_D=\log(2)$.  (d)  $S_{D,1/2}(\rho_\tGH)$ at $(h_x,h_z,J)=(0.26,0,0)$.  Inset: No correction to the area law is detected.}
    \label{EE_Z_2_GH}
\end{figure}

We now compare the above discussion of the usual EE, which ignores the exact Gauss law constraint, with the distillable EE. We focus on the distillable R\'enyi EE, which is easier to obtain from PEPS numerics than the von Neumann EE. While the definition of the distillable R\'enyi EE is not straightforward, a proper definition was provided in Ref.~\cite{Accessible_Renyi_entropy}. From the block diagonal structure of $\rho_{\tGH}$ shown in Eq.~\eqref{RDM_block_diagonal}, One can write the $n$-R\'enyi entropy as~\cite{Accessible_Renyi_entropy}:
\begin{equation}
S_{n}(\rho_{\tGH})=\frac{1}{1-1/n}\log\sum_{\pmb{x}}p_{\pmb{x},n}^{1/n}+S_{D,n}, \quad\quad
\end{equation}
where $p_{\pmb{x},n}=\Tr\bra{\pmb{x}}\rho_{\tGH}^{n}\ket{\pmb{x}}/\Tr\rho_{\tGH}^n$ and the term
\begin{equation}
   S_{D,n}=\frac{n}{1-n}\log\sum_{\pmb{x}}p_{\pmb{x}}\exp\left[\frac{1-n}{n}S_{n}(\rho_{\tGH,\pmb{x}})\right].
\end{equation}
is the distillable R\'enyi EE. When $n=1/2$, we have
\begin{equation}
    S_{D,1/2}=\log\sum_{\pmb{x}}p_{\pmb{x}}\left(\Tr\sqrt{\rho_{\tGH,\pmb{x}}}\right)^2.
\end{equation}
Because the reduced density matrices of the TC model in a field and the Fradkin-Shenker model are related by the quantum channel shown in Eq.~\eqref{eq:quantum_channel}, $S_{D,1/2}(\rho_{\tGH})$ and $S_{1/2}(\rho_{\tGH})$ of the Fradkin-Shenker model can be directly calculated from the transfer matrix fixed points of the PEPS $\ket{\Psi_{\tTC}}$ of the TC model in a field as shown in Appendix~\ref{App: Tensor_network_method_for_distillable_TEE}. The distillable EE $S_{D,1/2}$ is the sub-leading part of $S_{1/2}$, and is very small compared to $S_{1/2}$, so it is necessary to consider higher-order perturbations. However, since calculating higher-order perturbed ground state is cumbersome, we instead adopt the variational ground states in terms of PEPS~\cite{Corboz_2020} to numerically extract the distillable EE. We find that a very large circumference $N$ is needed to determine the behavior of $S_{D,1/2}$, and we derive a method to efficiently calculate $S_{D,1/2}$ for arbitrary $N$ in Appendix.~\ref{App: Tensor_network_method_for_distillable_TEE}. Here, we discuss the results. 

First, we investigate the EE of the pure $\mathbb{Z}_2$ lattice gauge theory, i.e., $h_z=J=0$, where the Hamiltonian has the exact 1-form 't Hooft loop symmetry.
We compute the total and distillable R\'enyi EE densities of $\ket{\Psi_{\tGH}}$ along the $h_x$ axis.
Our numerical results in Fig.~\ref{EE_Z_2_GH}a and b show that the distillable EE is non-zero for the pure $\mathbb{Z}_2$ lattice gauge theory with finite $h_x$. This is in contrast with Ref.~\cite{Distillable_entanglement_2016} which conjectures that the distillable EE is zero for the pure $\mathbb{Z}_2$ lattice gauge theory; we provide analytical analysis for our findings in Sec.~\ref{sec:distillable_ent_non_zero}.

Next, we consider how the distillable EE depends on the length $N$ of the entanglement cut. In Fig.~\ref{EE_Z_2_GH}c, we find that the distillable R\'enyi EE in the deconfined phase with the emergent 1-form 't Hooft loop symmetry is ($h_z=0,J\neq0$) contains the TEE, i.e., $S_{D,1/2}=\alpha N-\log(2)$, which is consistent with the perturbative calculation in Ref.~\cite{Distillable_entanglement_2016}.
Surprisingly, we find that the distillable EE in the deconfined phase with the exact 1-form 't Hooft loop symmetry ($h_z=0$) satisfies the area law \emph{without} TEE,  i.e., $S_{D,1/2}=\alpha N$, as shown in Fig.~\ref{EE_Z_2_GH}d. We now turn to an explanation of this phenomenon based on higher-form symmetries. 
\begin{figure}
    \centering
    \includegraphics{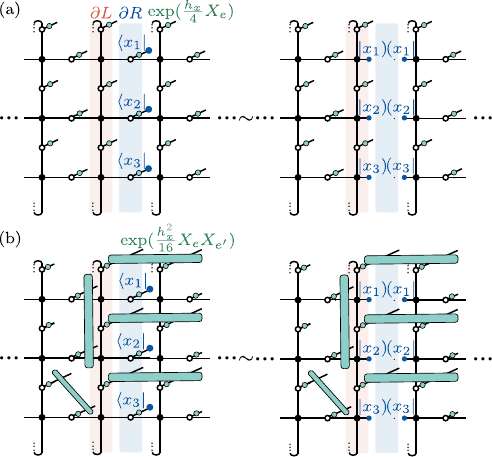}
    \caption{\textbf{Analyzing the distillable EE of the pure $\mathbb{Z}_2$ lattice gauge theory.} (a) Measuring the physical degrees of freedom of the first-order exponentiated perturbed PEPS factorizes it into two parts, giving zero distillable EE. (b) Measuring the physical degrees of freedom of the second-order exponentiated perturbed PEPS (many other two-site gates applying within the plaquettes are not shown for simplicity) does \emph{not} factorize it into two parts, which implies that the distillable EE is non-zero in general.}
    \label{Non-zero_distillable_TEE}
\end{figure}

\subsubsection{Explaining the absence of TEE in the distillable EE when the 1-form 't Hooft loop symmetry is exact}

Let us analyze the numerical results more carefully. The distillable von Neumann EE is a weighted average over many pure measured states $\ket{\Psi_{\tGH,\pmb{x}}}$, as shown Eq.~\eqref{distillable_von_Neumann_TEE}. We thus consider the EE of a single measured state $\ket{\Psi_{\tGH,\pmb{x}}}$ of the Fradkin-Shenker model, which is the same as the EE of the corresponding measured state  $\ket{\Psi_{\tTC,\pmb{x}}}=P_{\pmb{x}}\ket{\Psi_{\tTC}}/\sqrt{p_{\pmb{x}}}$ of the TC model in a field. This is because the measurement at $\partial R$ reduces the quantum channel to an isometry transformation relating $\rho_{\tTC,\pmb{x}}$ and $\rho_{\tGH,\pmb{x}}$ such that $S(\rho_{\tTC,\pmb{x}})=S(\rho_{\tGH,\pmb{x}})$, see details in Appendix~\ref{sec:ES_EH_IGH}.
As analyzed before, spontaneous 1-form symmetry breaking is a necessary condition for TEE. We should therefore check whether  $\ket{\Psi_{\tTC,\pmb{x}}}$ breaks the 1-form symmetries.

Recall that the fixed point ground state $\ket{\tTC}$ in Eq.~\eqref{fixed_point_TC} is an eigenstate of the non-contractible Wilson loop operator but not an eigenstate of the non-contractible 't Hooft loop operator. Let us consider a ground state $\ket{\Psi_{\tTC}}$ that is adiabatically connect with $\ket{\tTC}$ and thus share the same eigenvalue structure of the non-contractible Wilson and the 't Hooft loop operators. 
Suppose that the reduced density matrix of $\ket{\Psi_{\tTC}}$ is $\rho_{\tTC}\sim\exp(-H_E)$, where the EH $H_E$ has the local. Then, the EH of $\ket{\Psi_{\tTC,\pmb{x}}}$ is also local:
\begin{equation}
   \rho_{\tTC,\pmb{x}}\propto \prod_{e\in\partial R}\frac{1+x_eX_e}{2}\exp(-H_E)\propto\lim_{\beta\rightarrow+\infty}\exp(\beta \sum_e x_eX_e-H_E). \notag
\end{equation}
Next we just need to check if $\rho_{\tTC,\pmb{x}}\propto T_{\hat{C}_y}\rho_{\tTC,\pmb{x}}$ to determine whether TEE can be extracted from the measured states shown in Fig.~\ref{figure_distillable}b.

We at first consider the pure $\mathbb{Z}_2$ lattice gauge theory with $h_z=J=0$
, where the exact 1-form 't Hooft loop symmetry breaks spontaneously in the deconfined phase. However,
the measured state becomes an eigenstate of $T_{\hat{C}_y}$, as shown in Fig.~\ref{figure_distillable}b, so we have $T_{\hat{C}_y}\rho_{\tTC,\pmb{x}}=\left(\prod_{e\in \partial R}x_e\right)\rho_{\tTC,\pmb{x}}$,
which means that the exact 1-form 't Hooft loop symmetry $T_{\hat{C}_y}$ does not break. 
By our previous derivation, $(1+T_{\hat{C}_y})\ket{\Psi_{\tTC,\pmb{x}}}/2$ does not contain the TEE: $S_1(\rho_{\tTC,\pmb{x}})=\alpha_{\pmb{x}}N$  for sufficiently large $N$ if $\prod_{e\in \partial R}x_e=1$, where $\alpha_{\pmb{x}}$ is a non-universal coefficient. Notice that $S_1(\rho_{\tTC,\pmb{x}})=0$ if $\prod_{e\in \partial R}x_e=-1$.  
Taking the relation $S_1(\rho_{\tGH,\pmb{x}})=S_1(\rho_{\tTC,\pmb{x}})$ into consideration, the distillable von Neumann EE $S_{D,1}$ also satisfies an area law without the TEE: $S_{D,1}=N\left(\sum_{\pmb{x}}p_{\pmb{x}}\alpha_{\pmb{x}}\right)$, for sufficiently large $N$.

On the other hand, when the 1-form 't Hooft loop symmetry is an emergent symmetry ($h_z=0,J\neq0$), we can derive a unitary transformation $U$ relating the fixed point ground state $\ket{\tTC}$ and the first order perturbed ground state $\ket{\Psi_{\tTC}}$: $\ket{\Psi_{\tTC}}=\left[U+O(\delta'^2)\right]\ket{\tTC}$, from which we know that the emergent 1-form 't Hooft loop symmetry is $\tilde{T}_{\hat{C}}=UT_{\hat{C}}U^{\dagger}$. Then we can show that the measured state $\ket{\Psi_{\tTC,\pmb{x}}}$ still spontaneously breaks the emergent 1-form 't Hooft symmetry $\tilde{T}_{\hat{C}_y}$ on a non-contractible loop such that $\tilde{T}_{\hat{C}_y}\rho_{\tTC,\pmb{x}}\not\propto \rho_{\tTC,\pmb{x}}$, 
see details in Appendix~\ref{App:explicit_form_of_emergent_1-form_sym}. According to the previous derivations, 
the TEE can be extracted from the EE of each measured state: $S_1(\rho_{\tTC,\pmb{x}})=\alpha_{\pmb{x}}N-\log2$.
Taking the relation $S_1(\rho_{\tGH,\pmb{x}})=S_1(\rho_{\tTC,\pmb{x}})$ into account, the distillable von Neumann EE also contains the TEE:  $S_{1,D}=\sum_{\pmb{x}}\left(p_{\pmb{x}}\alpha_{\pmb{x}}\right)N-\log2$. In summary, TEE is robust against measuring $X$ operators along the entanglement cut when the 1-form 't Hooft loop symmetry is emergent, but it is fragile when the the 1-form 't Hooft loop symmetry is exact.

\subsubsection{Non-zero distillable EE for the pure $\mathbb{Z}_2$ lattice gauge theory}\label{sec:distillable_ent_non_zero}

 Previous work~\cite{Distillable_entanglement_2016} conjectured that the distillable EE is zero for the pure $\mathbb{Z}_2$ lattice gauge theory ($h_z=J=0$). Contrary to this, in our numerical calculation, we find that it can be non-zero. In order to understand this, we compute a second-order exponentiated perturbed wavefunction~\cite{Laurens_bridge_2017}:
\begin{equation}\label{variational_ansatz}
|\Psi^{(2)}_{\tTC}(h_x,0)\rangle=\exp[\sum_e \frac{h_x}{4} X_e+\sum_{\langle e e'\rangle\in p}
\frac{h_x^2}{16} X_e X_{e'}] \ket{\text{TC}},
\end{equation}
where we ignore the terms in $O(h_x^3)$. The difference between the second-order exponentiated perturbed wavefunction and the usual second-order perturbed wavefunction is $O(h_x^3)$. Notice that when $h_z=J=0$, the TC model in a field and the Fradkin-Shenker model are equivalent up to a product state, so we use the ground state of the TC model in a field. The reason that we consider the exponentiated perturbed wavefunction instead of the usual perturbed wavefunction is that the former can be exactly expressed as a PEPS.
First, we will only keep the first order term:
\begin{equation}
\ket{\Psi^{(1)}_{\tTC}(h_x,0)}=\exp(\sum_e\frac{h_x}{4}X_e)\ket{\tTC}.
\end{equation}
After measuring the physical degrees of freedom in $\partial R$, one finds that the PEPS $\ket{\Psi^{(1)}_{\tTC}(h_x,0)}$ factorizes into disconnected parts, see Fig.~\ref{Non-zero_distillable_TEE}a. This gives $S_n(\rho_{\tGH,\pmb{x}})=S_n(\rho_{\tTC,\pmb{x}})=0, \forall\pmb{x}$ and the distillable EE $S_{D,n}$ is zero $\forall n$. In Ref.~\cite{Distillable_entanglement_2016} only this order perturbation is considered, from which it was conjectured that distillable EE is zero. However, we will see that non-zero contributions can be found from the second order perturbation.

Therefore now we consider $\ket{\Psi^{(2)}_{\tTC}(h_x,0)}$. As shown in Fig.~\ref{Non-zero_distillable_TEE}b, the terms with a coefficient $h_x^2/16$ in Eq.~\eqref{variational_ansatz} are two-site non-unitary gates applied within the same plaquette. After measuring the physical degrees of freedom in $\partial R$, it can be found that the PEPS $\ket{\Psi^{(2)}_{\tTC}(h_x,0)}$ is \emph{not} factorized since the two-site gates in Fig.~\ref{Non-zero_distillable_TEE}b crossing $\partial R$ connect left and right parts. Hence the EE  $S_n(\rho_{\tTC,\pmb{x}})$ of the pure state $\ket{\Psi_{\tTC,\pmb{x}}}$ as well as the distillable EE $S_{D,n}$ can be non-zero. By decomposing the two-site gate: $\exp(h_x^2 X_1X_2/16)=\cosh(h_x^2/16)+\sinh(h_x^2/16)X_1X_2$, we can estimate the  distillable 1/2-R\'enyi EE density as:
  \begin{equation}\label{decompose_gate}
S_{D,1/2}\approx 2\log(\sqrt{\frac{\cosh^2(h_x^2/16)}{\cosh(h_x^2/8)}}+\sqrt{\frac{\sinh^2(h_x^2/16)}{\cosh(h_x^2/8)}}),
  \end{equation}
  which is consistent with the results from the tensor network numerics, as shown in the inset of the Fig.~\ref{EE_Z_2_GH}b. From
Fig.~\ref{Non-zero_distillable_TEE}b, we conclude that although $X$ measurements along $\partial R$ destroy the underlying long-range entanglement of $\ket{\Psi_{\tTC,\pmb{x}}}$ along the entanglement cut, some short-range entanglement can be retained by the two-site non-unitary gates (and more complicated gates from higher order perturbations), leading to non-zero distillable EE without TEE when tuning $h_x$ at fixed $J=0$ and $h_z=0$.

\section{Discussion and outlook}\label{discussion_and_outlook}

In this paper, we investigated the entanglement properties of the gapped phases of an emergent $\mathbb{Z}_2$ lattice gauge theory from the perspective of 1-form symmetries. We find that the entanglement properties of the model can be understood from whether the 1-form symmetries are exact or emergent and whether the Gauss law is exact or emergent. Therefore, our work demonstrates that higher-form symmetries provide a very useful tool for understanding entanglement properties of exact and emergent gauge theories.

First, we found that in the non-trivial SPT phase, the entanglement spectrum (ES) generally has a 2-fold degeneracy, which is robust against the perturbations that explicitly break the exact 1-form Wilson loop symmetry. Furthermore, when both the 1-form Wilson loop symmetry and the Gauss law constraint is exact, the ES degeneracy is enhanced to $2^N$. 
However, in the Higgs regime of the Fradkin-Shenker model, with an exact Gauss law constraint and an emergent 1-form Wilson loop symmetry, the ES is generally non-degenerate. Comparing these results to those on the physical boundaries in Ref.~\cite{Ruben_Higgs_SPT}, we find that the physical boundary needs to be fine-tuned to realize the bulk-boundary correspondence.

 Second, we consider the ES in the topologically ordered phase where 1-form symmetries are spontaneously broken. Because the fixed point ground states of the topologically ordered phase and that of the non-trivial SPT phase are related via an global on-site projector, we find that the EH near the fixed point of this topologically ordered phase has similar terms to the one near the fixed-point limit of the non-trivial phase. However, the EH in the topologically ordered phase also contains a non-local projector which stems from 1-form symmetry breaking and removes half of the spectrum, leading to a non-degenerate ES in the topologically ordered phase in general.

Third, we constructed a direct connection between spontaneous higher-form symmetry breaking and topologial entanglement entropy (TEE), which can be used to understand whether the distillable EE of the Fradkin-Shenker model with the exact Gauss law constraint contains the TEE. From the fact that 1-form symmetry breaking is necessary for the presence of the TEE, we were able to demonstrate that TEE is robust against the measurement when the 1-form symmetry is emergent  but it is fragile when the 1-form symmetry is exact. Therefore, when the 1-form 't Hooft loop symmetry is exact, the TEE is missing from the distillable EE even in the deconfined phase. Our results imply that when the Gauss law constraint is exact, the entanglement properties are qualitatively different, depending on whether the higher-form symmetries are exact or emergent. Our results can be straightforwardly generalized to the other lattice gauge theories with discrete Abelian symmetries.

Our results lead to a number of open questions deserving further exploration. For instance, it is interesting to explore the full phase diagram of the emergent $\mathbb{Z}_2$ lattice gauge theory out of the three planes in Fig.~\ref{Entanglement_cut}e. By analogy to quotient SPT phases~\cite{quotient_SPT_2021}, we expect that the phase transition between the trivial and non-trivial SPT phases is no longer necessary when the 1-form Wilson loop symmetry is not exact, and the four phases of the cluster model in a field may reduce to three phases in a $(J,h_z)$ plane with some finite $h_x$. Generalizing our results to gauge theories with gauge groups other than $\mathbb{Z}_2$ is another avenue for further study. 

Another intriguing aspect of our findings is the existence of a stable regime where the lowest-lying entanglement eigenvalues are degenerate, even when the higher-form symmetry is broken and thus the state is no longer an SPT in the standard sense. The relationship of this phenomenon with emergent higher-form symmetries~\cite{Wen_emergent_high_form_2023} and measurement-based quantum computation~\cite{raussendorf2005long} warrants further study. Surprisingly, we found that the Lieb-lattice cluster state, which serves as the obvious fixed-point Hamiltonian for this phase, actually sits at the boundary of this stable regime, due to its enhanced $2^N$-fold degeneracy. It remains an open question whether there exists a zero-correlation length fixed point ground state for this SPT phase which only has a $2$-fold degeneracy that is robust to 1-form symmetry breaking perturbations.

 The non-trivial SPT protected by the 1-form and the 0-form $\mathbb{Z}_2$ symmetries is analogy to the 1-dimensional non-trivial SPT protected by a 0-form $\mathbb{Z}_2\times\mathbb{Z}_2$ symmetry, in the sense that they share similar projective representations at the entanglement degrees of freedom, which lead to a 2-fold necessary degeneracy in the ES. This connection deserves further investigation which could be helpful for classifying such higher-form SPT phases.

On a more technical side, it would be useful to generalize the isometry transformation and the quantum channel between the Fradkin-Shenker model and TC model in a field to other gauge theories~\cite{bond_algenra_2011}, such that we can investigate the entanglement of gauge theories without considering the Gauss law constraint and matter field. 
If such a generalization exists, the PEPS of gauge theories~\cite{Gauge_TNS_continuous_2014,Gauging_2015} can be simplified, and it becomes easier to investigate their entanglement properties both analytically and numerically.

\textbf{Acknowledgements.} We especially thank Ari Turner for many useful discussions on this work and prior collaborations on entanglement transitions. We also thank R.-Z. Huang and Ruben Verresen for helpful comments. TR thanks Vedika Khemani, Yimu Bao, Curt von Keyserlingk and especially Yaodong Li for useful discussions. We acknowledge support from the Deutsche Forschungsgemeinschaft (DFG, German Research Foundation) under Germany’s Excellence Strategy--EXC--2111--390814868, TRR 360 – 492547816 and DFG grants No. KN1254/1-2, KN1254/2-1, the European Research Council (ERC) under the European Union’s Horizon 2020 research and innovation programme (grant agreement No. 851161 and No. 771537), as well as the Munich Quantum Valley, which is supported by the Bavarian state government with funds from the Hightech Agenda Bayern Plus. T.R. is supported in part by the Stanford Q-Farm Bloch Postdoctoral Fellowship in Quantum Science and Engineering.

\textbf{Data availability --} Data, data analysis, and simulation codes are available upon reasonable request on Zenodo~\cite{this_paper}.

\appendix

\section{Deriving entanglement Hamiltonians}\label{Appendix_EH}

In this section, we derive the EHs by combining the fixed point PEPS $\ket{\tcl}$ at $(h_x,h_z,J)=(0,0,\infty)$ and $\ket{\tTC}$ at $(h_x,h_z,J)=(0,0,0)$ and perturbation theory, separately. 

The reduced density matrix  $\rho$ of an injective PEPS $\ket{\Psi}$ contains a large null space, an isometric transformation $\mathcal{V}$ can be found to transform $\rho$ to the basis spanned by eigenvectors of $\rho$ with non-zero eigenvalues. The transformed reduced density matrix can be expressed in terms of left and right fixed points $\sigma_L$ and $\sigma_R$ of the PEPS transfer matrix~\cite{Cirac_2011}:
  \begin{equation}\label{rho_and_sigma}
\mathcal{V}\rho \mathcal{V}^{\dagger}=\mathcal{V}\left(\Tr_L\ket{\Psi}\bra{\Psi}\right)\mathcal{V}^{\dagger}\propto\sqrt{\sigma_R^{\text{T}}}\sigma_L\sqrt{\sigma_R^{\text{T}}},
\end{equation}
where the isometry $\mathcal{V}$  satisfies $\mathcal{V}\mathcal{V}^{\dagger}=\mathbbm{1}$ and $\mathcal{V}^{\dagger}\mathcal{V}$ satisfying $[\mathcal{V}^{\dagger}\mathcal{V},\rho]=0$ is a projector which projects out the null space of $\rho$.

\subsection{EH near the fixed point of the topologically ordered phase}\label{App_A_1}

\begin{figure}
    \centering
    \includegraphics{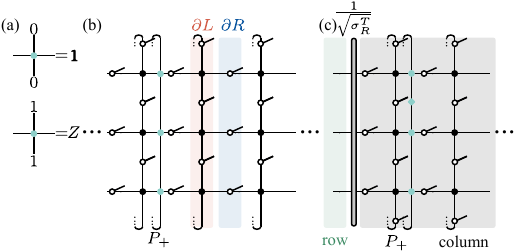}
    \caption{\textbf{The PEPS of the trivial MES and the isometry transformations to discard the null space.} (a) The tensors define $P_{+}=(1+\prod_iX_i)/2$. (b) The PEPS for the trivial MES. (c) The tensor network representation of the isometry $\mathcal{V}$, the dangling virtual legs on the left side consist of the row index of the isometry, and the physical legs on the right part of the system consist of the column index of the isometry. At the fixed point of the topologically ordered phase, we can simply take $\sigma_R=\mathbbm{1}$.}
    \label{Figure apppendix_4}
\end{figure}

We derive the EH by perturbing the fixed point trivial minimally entangled state (MES) $(1+T_{C_y})\ket{\tTC}/\sqrt{2}$ at $(h_x,h_z,J)=(0,0,0)$, as shown in Figs.~\ref{Figure apppendix_4}a and b (ignoring the product state $\prod_v\ket{+}_v$). From Ref.~\cite{Schuch_transfer_operator_2013}, we know that the transfer matrix of the PEPS $\ket{\tTC}$ can be expressed as $\mathbb{T}=\mathbbm{1}_2^{\otimes N}\otimes\mathbbm{1}_2^{\otimes N}+X^{\otimes N}\otimes X^{\otimes N}$. Since the PEPS $\ket{\tTC}$ is $\mathbb{Z}_2$ injective, the left and right fixed points of the transfer matrix are two-fold degenerate: $\sigma_L=\sigma_R=\mathbbm{1}_2^{\otimes N}$ or $X^{\otimes N}$. The effective reduced density matirx can be constructed from the superpositions of the degenerate transfer matrix fixed points~\cite{Schuch_transfer_operator_2013}:
\begin{equation}
    \mathcal{V}\rho^{(0)}_{\tTC}\mathcal{V}^{\dagger}=\frac{1}{2^{N}}(\mathbbm{1}_2^{\otimes N}+X^{\otimes N})=\frac{1}{2^{N-1}}P_+.
\end{equation}
 where, $\mathcal{V}$ defined in Figs.~\ref{Figure apppendix_4}c, is simply half of the PEPS, and it is an isometry when restricted to the virtual subspace with even parity.

Next, let us consider the first-order perturbed wavefunction shown in Eq.~\eqref{eq:perturbed_MES}.
The reduced density matrix can be written as
\begin{align}\label{RDM_TC_1}
    \rho_{\ttop}&=\Bigg[\Tr_{L} \ket{\text{TC}}\bra{\text{TC}}+\frac{h_x}{4}\Tr_{L}\sum_{e\in E}\left(X_e\ket{\text{TC}}\bra{\text{TC}}+\text{h.c.}\right)\notag\\
&+\frac{h_z}{4}\Tr_{L}\sum_{e \in E}\left(Z_e\ket{\text{TC}}\bra{\text{TC}}+\text{h.c.}\right)\notag\\
 &+\frac{J}{4}\Tr_{L}\sum_{\langle vev'\rangle}\left(Z_vZ_eZ_{v'}\ket{\text{TC}}\bra{\text{TC}}+\text{h.c.}\right)+O(\delta_r^2)\Bigg]\notag\\
 & \frac{1+T_{\hat{C}_y}}{2}\otimes\prod_{v\in R}\frac{1+X_v}{2}.
\end{align}
Here, we also need an isometry to project onto the low-entanglement energy subspace. We can use a modified isometry $\mathcal{V}'=\mathcal{V}\otimes\prod_{v\in R}\bra{+}_v$ to take the vertex degrees of freedom into account. From the Schrieffer–Wolff transformation~\cite{S_W_transformation_2011}, we know that the error of using $\mathcal{V}'$ as an approximate isometry is $O(\delta_r^2)$, so we can safely obtain the effective reduced density matrix via:
\begin{align}\label{V_rho_V}
   &\mathcal{V}'\rho_{\ttop} \mathcal{V}^{\dagger}=[\mathcal{V}\Tr_{L} \ket{\text{TC}}\bra{\text{TC}} \mathcal{V}^{\dagger}\notag\\
    &+\frac{h_x}{4} \mathcal{V}\Tr_{L}\sum_{e\in \partial R}\left(X_e\ket{\text{TC}}\bra{\text{TC}}+\text{h.c.}\right) \mathcal{V}'^{\dagger}\notag\\&+ \frac{h_z}{4}\mathcal{V}\Tr_{L}\sum_{e\in\partial L}\left(Z_e\ket{\text{TC}}\bra{\text{TC}}+\text{h.c.}\right)]\mathcal{V}^{\dagger}+O(\delta_r^2)],
\end{align}
where we used the relation:
\begin{equation}
    \mathcal{V}'\left(\Tr_{L}\sum_{\langle vev'\rangle}Z_vZ_eZ_{v'}\ket{\tTC}\bra{\tTC}\otimes\prod_v\frac{1+X_v}{2}\right)\mathcal{V}'=0,
\end{equation}
because in the Hamiltonian in Eq.~\eqref{EGT} there always exists terms $X_v$  which anticommutes with $Z_vZ_eZ_{v'}$.
Notice that in the last two terms, we only sum over edges in $\partial L$ or $\partial R$ because other terms do not contribute to the zeroth and the first orders. For example,  $\mathcal{V}\Tr_L\left( X_e\ket{\text{TC}}\bra{\text{TC}}\right)\mathcal{V}^{\dagger}=0,\forall e\in (E\setminus\partial R)$, because $\mathcal{V} B_p=\mathcal{V}, B_p\ket{\text{TC}}=\ket{\text{TC}}$ and $\{B_p,X_e\}=0, e\in p$, where $E$ is a set of all edges. For the same reason,
$\mathcal{V}\Tr_L\left(Z_e\ket{\tTC}\bra{\tTC}\right)\mathcal{V}^{\dagger}=0,\forall e\in (E\setminus\partial L)$.

Using the relation shown in Figs.~\ref{Fig_EH_TC}c and d, the terms of the effective EHs can be derived:
\begin{align}\label{promote_X_Z}
   \mathcal{V}\Tr_L\left( Z_e\ket{\text{TC}}\bra{\text{TC}}\right)\mathcal{V}^{\dagger}&=Z_iZ_{i+1}P_{+}/2^{N-1},  \notag\\
       \mathcal{V}\Tr_L\left( X_e\ket{\text{TC}}\bra{\text{TC}}\right)\mathcal{V}^{\dagger}&=X_iP_{+}/2^{N-1}.
\end{align}
Therefore, by substituting Eq.~\eqref{promote_X_Z} into Eq.~\eqref{V_rho_V}, we obtain the effective reduced density matrix of the trivial MES. The effective EH of the trivial MES is $H^{\text{eff}}_{E,\ttop}=-\log(\mathcal{V}\rho_{\ttop}\mathcal{V}^{\dagger})$, which can be simplified to the Eq.~\eqref{ent_ham_all} using the relation $\log(1+x)=x+O(x^2)$.
The method we use for deriving the EH is similar to that shown in Ref.~\cite{Yang_edge_2014}.

\subsection{EH near the fixed point of the non-trivial SPT phase}\label{EH_of_cluster_model}
The EH near the fixed point of the non-trivial SPT phase can be derived similarly. An isometry $\mathcal{V}_{\tcl}$ transforming $\rho_{\tcl}=\Tr_L\ket{\Psi_{\tcl}}\bra{\Psi_{\tcl}}$ to the effective reduced density matrix $\mathcal{V}_{\tcl}\rho_{\tcl}\mathcal{V}_{\tcl}^{\dagger}$ can be constructed similarly to that of $\rho_{\ttop}$ shown in Figs.~\ref{Figure apppendix_4}c, and $\mathcal{V}_{\tcl}$ is nothing but the $R$ part of the PEPS $\ket{\tcl}$ displayed in Fig.~\ref{Figure_Higgs_TN}b, because the left and right transfer matrix fixed points of $\ket{\tcl}$ are the identity matrix $\sigma_L=\sigma_R=\mathbbm{1}_{2^N}$.

Considering the perturbed cluster state in Eq.~\eqref{perturbed_cluster}, it can be shown that only terms $\sum_{e\in\partial R}X_e$ and  $\sum_{e\in\partial R}Z_e$ contribute to the first order of the EH similar to the case of the previous subsection. The term $\sum_vX_v$ does not contribute to the zeroth and first orders of the EH:
\begin{equation}
    \mathcal{V}_{\tcl}\Tr_{L}\left(\sum_v  X_v\ket{\tcl}\bra{\tcl}\right)\mathcal{V}^{\dagger}_{\tcl}=0,
\end{equation}
because in the Hamiltonian in Eq.~\eqref{Ham_cluster} there always exists at least one term $Z_vZ_eZ_{v'}$  which does not cross the entanglement cut and anticommutes with $X_v$. So, using the relation in Figs.~\ref{Figure_Higgs_TN}c and d and the fact that $\sigma_L=\sigma_R=\mathbbm{1}_{2^N}$, one can derive the EH in Eq.~\eqref{EH_SPT}.

\subsection{Low energy ES of the Fradkin-Shenker model and probability distribution of measurement outcomes}\label{EH_IGH_app}

From Eqs.~\eqref{EH_GH_trivial} and ~\eqref{EH:IGH_in_topo}, one finds that the dominant parts of the EHs of the Fradkin-Shenker model are classical.
Actually, they are related to the probability distribution of sub-blocks of the reduced density matrix. Considering the trivial MES sector, because $\mathcal{V}'X_{e\in\partial R}\mathcal{V}'=X_i$, the eigenstates of the $H^{\text{eff}}_{E,\tGH}$ can be labeled by $\pmb{x}=\{x_e|e\in\partial R\}$, and the ES can be expressed as $\epsilon_{\pmb{x}}=(N-1)\log 2-h_x\sum_{e\in \partial R}x_e/2+O(\delta'^2)$. Comparing with the probability distribution:
\begin{equation}
    p_{\pmb{x}}=\bra{\Psi_{\tGH}}P_{\pmb{x}}\ket{\Psi_{\tGH}}=\frac{1}{2^{N-1}}\left(1+\sum_e\frac{x_e h_x}{2}\right)+O(\delta'^2),
\end{equation}
we have $p_{\pmb{x}}=\exp(-\epsilon_{\pmb{x}})+O(\delta'^2)$. This relation is also valid in the Higgs-confined phase.

\section{Why is the ES of the Fradkin-Shenker model not symmetric about $\theta'=\pi/4$?}\label{app:why_ES_of_FS_non_symmetric}
We observed that the bulk spectrum of the Fradkin-Shenker model is symmetric about $\theta'=\pi/4$, but its ES is not. This raises the question of why the electric-magnetic duality transformation fails to enforce the ES of the Fradkin-Shenker model to be symmetric about $\theta'=\pi/4$. First, since the TC model in a field has the electric-magnetic duality symmetry: $U_{\tTC}H_{\tTC}(h_x,h_z)U^{\dagger}_{\tTC}=H_{\tTC}(h_z,h_x)$, where $U_{\tTC}$ is a unitary transformation which exchanges the primal and dual lattices as well as Pauli $X$ and $Z$ operators, the Fradkin-Shenker model also has a electric-magnetic duality symmetry: $U_{\tGH}H_{\tGH}(h_x,J)U^{\dagger}_{\tGH}=H_{\tTC}(J,h_x)$, where
$U_{\tGH}=V_{\tCX}U_{\tTC}V^{\dagger}_{\tCX}$. This is the reason that the bulk spectrum of the Fradkin-Shenker model is symmetric about $\theta'=\pi/4$.

 $U_{\tTC}$ induces a unitary electric-magnetic duality transformation $U_{E,\tTC}$ that is applied to $\rho_{\tTC}$: $\rho_{\tTC}(h_z,h_x)=U_{E,\text{TC}}\rho_{\tTC}(h_x,h_z)U^{\dagger}_{E,\text{TC}}$. On the other hand, the electric-magnetic duality transformation applied to $\rho_{\tGH}=\mathscr{N}[\rho_{\tTC}]$ is a completely positive map instead of a unitary transformation:
\begin{equation}
   \mathscr{U}_{E,\tGH}[\cdot]=\mathscr{N}[U_{E,\tTC}\mathscr{N}^{-1}[\cdot]U_{E,\tTC}^{\dagger}],
\end{equation}
where $\mathscr{N}^{-1}[\cdot]=\sum_{\pmb{z}}K^{\dagger}_{\pmb{z}}\cdot K_{\pmb{z}}$. 
Since $\rho_{\tGH}(J,h_x)=\mathscr{U}_{E,\tGH}[\rho_{\tGH}(h_x,J)]$ is no longer a unitary transformation, the spectra of $\rho_{\tGH}(h_x,J)$ and $\rho_{\tGH}(J,h_x)$ are generically different. So, using the quantum channel, we can explain the reason that the ES of the Fradkin-Shenker model is non-symmetric about $\theta'=\pi/4$ even if the Fradkin-Shenker model itself has the electric-magnetic duality symmetry.

\section{Tensor network method for calculating ES}\label{Appendix_iPEPS}

\begin{figure*}
    \centering
    \includegraphics{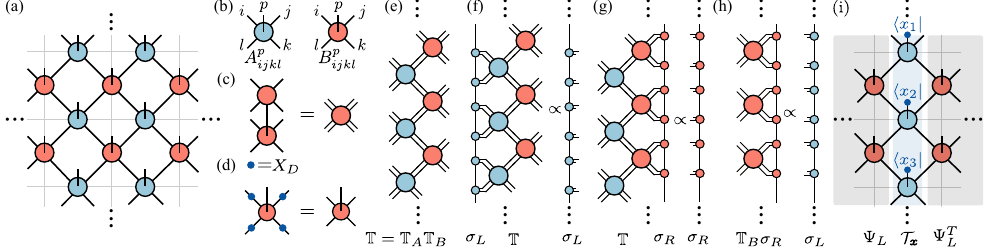}
    \caption{\textbf{The infinite PEPS ansatz, the transfer matrix, and the fixed points.} (a) The infinite PEPS ansatz for variational optimization. (b) The two tensors of the infinite PEPS are related by a reflection. (c) The double tensor of $A$. The double tensor of $B$ is similar. (d) The virtual $\mathbb{Z}_2$ symmetry of the tensor, there $X_D$ is a $D\times D$ matrix satisfying $X_D^2=1$. (e) The transfer matrix of the infinite PEPS. (f) The fixed point equation for the left fixed point $\sigma_L$.  (g) The fixed point equation for the right fixed point $\sigma_R$. (h) The relation between left and right fixed points. (i) The PEPS $\ket{\Psi_{\tTC}}$ can be viewed as a product of three matrices $\Psi_L$, $\mathcal{T}_{\pmb{x}}$ and  $\Psi_L^T$. }
    \label{Fig_append_1}
\end{figure*}

To calculate the effective reduced density matrix, it is convenient to express the ground state in terms of PEPS shown in Fig.~\ref{Fig_append_1}a generated by two rank-5 tensors with the virtual bond dimension $D$ and the physical dimension $d=2$, see Fig.~\ref{Fig_append_1}b. We can obtain the PEPS either variationally or perturbatively. For the toric code model in a field, we can use the variational infinite PEPS method introduced in Ref.~\cite{Corboz_2020}. Moreover, if we are close to the fixed point of gapped phases,  we can express the first-order perturbed ground state in terms of PEPS by exponentiating it~\cite{Laurens_bridge_2017}, i.e., re-express perturbed wavefunction $(1+\epsilon V)\ket{\Psi_0}=\left[\exp(\epsilon V)+O(\epsilon^2)\right]\ket{\Psi_0}$, where $\ket{\Psi_0}$ is the unperturbed wavefunction and $\epsilon V$ is a small perturbation. 
In the topological phase, the PEPS tensor has the symmetry shown in Fig.~\ref{Fig_append_1}d. It is convenient to calculate the effective reduced density matrix from the fixed points of the PEPS transfer matrix, as shown in Fig.~\ref{Fig_append_1}e. The transfer matrix $\mathbb{T}$ can be decomposed as a product of $\mathbb{T}_A$ and $\mathbb{T}_B$. Considering the shape of the transfer matrix, we use the iTEBD (infinite time evolution block decimation) algorithm to approximate the left fixed point $\sigma_L$ and the right fixed point $\sigma_R$ in terms of infinite MPS, see Figs.~\ref{Fig_append_1}f and \ref{Fig_append_1}g. Since $A$ and $B$ tensors have reflection symmetry, $\sigma_L$ and $\sigma_R$ are related, as shown in Fig.~\ref{Fig_append_1}h.  We can calculate the spectrum of the reduced density matrix using $\sqrt{\sigma^{\text{T}}_L}\sigma_R\sqrt{\sigma^{\text{T}}_L}\sim \sigma^{\text{T}}_L\sigma_R$.

\subsection{Near the product state limit}

We now derive a first order perturbed ground state in the trivial phase of the TC model in a field by perturbing from the $\delta=\sqrt{h_x^2+h_z^2}\rightarrow\infty$ limit.
We at first apply a unitary transformation
\begin{equation}
    U=\left(\begin{array}{cc}
       \cos (\theta/2)  & \sin (\theta/2)  \\
        -\sin (\theta/2) & \cos (\theta/2)
    \end{array}\right)
\end{equation}
 to the toric code mode in a field:
\begin{equation}
    U^{\dagger}H_{\tTC}U=-\sum_v A_v'-\sum_p B_p'-r\sum_e X_e,
\end{equation}
where $A'_v=\prod_{e\in v} X_e'$, $B'_p=\prod_{e\in p} Z_e'$, $X'=X\cos\theta -Z\sin \theta$ and $Z'=Z\cos\theta +X\sin \theta$. We first calculate in the $U$-transformed basis and then transform back to the original basis.
When $\delta\rightarrow{+\infty}$, the ground state of $H_{\tTC}$ can be written as
\begin{equation}
    \ket{\Psi^{(0)}}=\prod_e\ket{\theta}_e,\quad\ket{\theta}=U\ket{+}=\cos\frac{\theta}{2}\ket{+}+\sin\frac{\theta}{2}\ket{-}.
\end{equation}
When the field is large, the first order ground state can be written as
\begin{align}
    \ket{\Psi^{(1)}}&=\prod_e U_e\Biggl(1+\sum_v \prod_{e\in v}\sum_{\{\alpha_e=0,1\}}f_v(\{\alpha_e\},\delta,\theta)Z^{\alpha_i}_e\notag\\
    &+\sum_p \prod_{e\in p}\sum_{\{\alpha_e=0,1\}}f_p(\{\alpha_e\},\delta,\theta)Z^{\alpha_i}_e\Biggr)\prod_e\ket{+}_e,
\end{align}
where
\begin{align}
    f_v(\{\alpha_e\},\delta,\theta)&=\begin{cases}
     \frac{\prod_{e=1}^4(\cos\theta)^{1-\alpha_e}(-\sin\theta)^{\alpha_e}}{2\delta\sum_{e=1}^4\alpha_e},&\quad\mbox{if } \sum_{e=1}^4\alpha_e\neq 0;\\
     0,&\quad \mbox{else if }\sum_{e=1}^4\alpha_e=0;
    \end{cases}\\
     f_p(\{\alpha_e\},\delta,\theta)&=\begin{cases}
     \frac{\prod_{e=1}^4(\sin\theta)^{1-\alpha_e}(\cos\theta)^{\alpha_e}}{2\delta\sum_{e=1}^4\alpha_e},&\quad\mbox{if } \sum_{e=1}^4\alpha_e\neq 0;\\
     0,& \quad \mbox{else if }\sum_{e=1}^4\alpha_e= 0.
      \end{cases}
\end{align}
We can exponentiate $\ket{\Psi^{(1)}}$ such that it can be written as a PEPS:
\begin{align}
    \label{eq:suppPEPSPert}\ket{\tilde{\Psi}^{(1)}}&=\prod_eU_e\Bigg\{\prod_v\exp\left[ \prod_{e\in v}\sum_{\{\alpha_e=0,1\}}f_v(\{\alpha_e\},\delta,\theta)Z^{\alpha_i}_e\right]\notag\\
    &\times\prod_p\exp\left[\prod_{e\in p}\sum_{\{\alpha_e=0,1\}}f_p(\{\alpha_e\},\delta,\theta)Z^{\alpha_i}_e\right]\Bigg\}\prod_e\ket{+}_e.
\end{align}
The difference between $\ket{\tilde{\Psi}^{(1)}}$ and  $\ket{\Psi^{(1)}}$ is $O(1/\delta^2)$.
Now, $\ket{\tilde{\Psi}^{(1)}}$ can be interpreted as some gates applied on a product state, and it can be written as a $2\times2$ unit cell PEPS shown in Fig.~\ref{Fig_append_1}a with bond dimension $4$.

\begin{figure}
    \centering
    \includegraphics[scale=0.5]{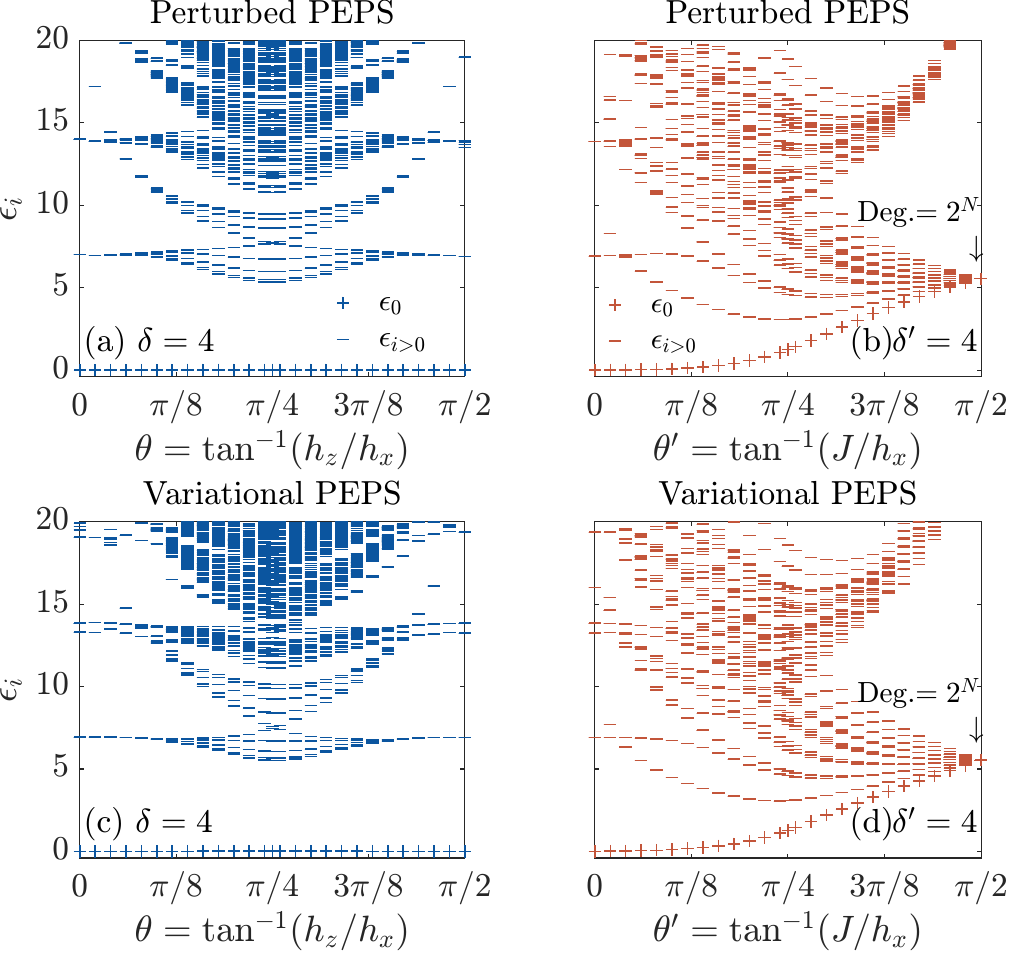}
    \caption{\textbf{Comparing the ES obtained from the variational PEPS and perturbed PEPS.} ES $\{\epsilon_i\}$ ($\epsilon_0\ge\epsilon_1\ge \epsilon_2\ge\cdots$) for systems on an infinitely long cylinder with a finite circumference $N=8$, where the fixed points of the PEPS transfer matrices are approximated by matrix product states with a dimension $20$. 
    (a) ES of the TC model in a field obtained from the exponentiated perturbed PEPS along the path $\delta=4$. (b)  ES of the Fradkin-Shenker model obtained from the exponentiated perturbed PEPS along the path $\delta'=4$. (c) ES of the TC model in a field obtained from the variational PEPS along the path $\delta=4$. (d)  ES of the Fradkin-Shenker model obtained from the variational PEPS along the path $\delta'=4$.}
    \label{Fig:app_compare}
\end{figure}

Alternatively, we can use the variational PEPS to calculate the ES. Figs.~\ref{Fig:app_compare}a and c show the ES calculated using the variational PEPS and the exponentiated perturbed PEPS at the field strength $\delta=4$ and the circumference $N=8$. We observe that the low energy parts of the ES from the two methods are almost identical, while there are some slight differences in the higher parts of the ES. So we just use the perturbed PEPS to calculate the low energy part of ES when close to some fixed points of gapped phases.

\subsection{Near the fixed point of the non-trivial SPT state}
Near the fixed point of the non-trivial SPT phase, the first order perturbed wavefunction in Eq.~\eqref{perturbed_cluster} can be exponentiated as
\begin{align}
    \ket{\Psi^{(1)}_{\tcl}}&=\prod_e\exp\left[\frac{\delta}{2}\left(\frac{\cos\phi
 Z_e}{2}+\sin\theta_x X_e\right)\right]\notag\\
&\times\prod_v\exp\left(\frac{\delta\cos\phi}{8}\sum_vX_v\right)\ket{\tcl}.
\end{align}
It can be easily written as a PEPS by just applying the product of local perturbations to the fixed point PEPS $\ket{\tcl}$ shown in Fig.~\ref{Figure_Higgs_TN}b, and we can use the standard method calculating ES of the exponentiated perturbed PEPS.

\subsection{Near fixed point of the topologically ordered phase}

Near the fixed point $(h_x,h_z,J)=(0,0,0)$ of the topologically ordered phase, the first order pertured ground state is obtained in Eq.~\eqref{eq:perturbed_MES}, which can be exponentiated as
\begin{align}\label{eq:exp_perturbed_MES}
  \ket{\Psi^{(1)}_{\ttop}}&=\prod_e\exp\left[
  \frac{h_x}{4}\sum_e X_e+\frac{h_z}{4}\sum_e Z_e+\frac{J}{4}\sum_{\langle vev'\rangle} Z_vZ_eZ_{v'}
  \right]\notag\\
  &\times\frac{1 +T_{\hat{C}_y}}{\sqrt{2}}\ket{\tTC}\otimes \prod_v\ket{+}_v,
\end{align}
up to an error $O(\delta_r^2)$. It can be easily written as a PEPS by simply applying the product of local perturbations to the fixed point PEPS $\ket{\tTC}$ shown in Fig.~\ref{Fig_EH_TC}b, and we can use the standard method for calculating the ES of the exponentiated perturbed PEPS. Notice that we can pull the MES projector \label{eq:exp_perturbed_MES} from the physical level to the virtual level, which becomes $P_{\pm}=(1\pm X_D^{\otimes N})/2$, where $X_D$ is the virtual symmetry of the PEPS in Fig.~\ref{Fig_append_1}d. Since the fixed points $\sigma_{L}$ and $\sigma_{R}$ of the transfer matrix inherit the virtual $\mathbb{Z}_2$ symmetry: $[\sigma_{L},X_D^{\otimes N}]=0$ and $[\sigma_{R},X_D^{\otimes N}]=0$.
 The ES of the trivial MES can be obtained from the relation:
\begin{equation}\label{eq_RDM_TC}
    \rho_{\ttop}\sim P_{+}\sigma_L^{\text{T}}\sigma_R/\Tr(P_{+}\sigma_L^{\text{T}}\sigma_R).
\end{equation}

\subsection{ES of the Fradkin-Shenker model}\label{sec:ES_EH_IGH}
In this section, we show how to calculate the ES of the subblock reduced density matrix $\rho_{\tGH,\pmb{x}}$ of the Fradkin-Shenker model from the transfer matrix fixed points of the corresponding PEPS of the TC model in a field. By measuring the physical degrees of freedom in $\partial R$, the quantum channel reduces to $VP_{\pmb{x}}\cdot P_{\pmb{x}}V^{\dagger}$:
\begin{equation}
  \vcenter{\hbox{
  \includegraphics{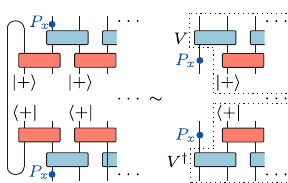}}}
\end{equation}
 where the isometry operator $V$ is defined in the box of dotted lines and applies on the $R$ part of the system. Therefore $\rho_{\tGH,\pmb{x}}=V\rho_{\tTC,\pmb{x}}V^{\dagger}$, and $\rho_{\tGH,\pmb{x}}$ and $\rho_{\tTC,\pmb{x}}$ have the same spectrum, where $\rho_{\tTC,\pmb{x}}$ is the reduced density matrix of $\ket{\Psi_{\tTC,\pmb{x}}}=P_{\pmb{x}}\ket{\Psi_{\tTC}}/\sqrt{p_{\pmb{x}}}$. As shown in Fig.~\ref{Fig_append_1}i, $\bra{\pmb{x}}\ket{\Psi_{\tTC}}$ can be decomposed into three parts:
\begin{equation}
\bra{\pmb{x}}\ket{\Psi_{\tTC}}=\Psi_L\mathcal{T}_{\pmb{x}}\Psi_L^{\text{T}},
\end{equation}
where $\Psi_L$ is a matrix whose row index is the collection of all physical indices in $L$ and column index is the collection of all virtual indices at the entanglement cut, and $\mathcal{T}_{\pmb{x}}$ is a matrix whose row (column) is a collection of left (right) virtual indices of $A$ tensors whose physical indices are fixed by the $X$ measurement. From Fig.~\ref{Fig_append_1}f, we have $\Tr_L\bra{\pmb{x}}\ket{\Psi_{\tTC}}\bra{\Psi_{\tTC}}\ket{\pmb{x}}=\Psi^{*}_L\mathcal{T}_{\pmb{x}}^{\dagger}\sigma_L \mathcal{T}_{\pmb{x}}\Psi_L^{\text{T}}$, because $\Psi_L^{\dagger}\Psi_L\propto\sigma_L$, where $\sigma_L$ is the left transfer matrix fixed point of the PEPS $\ket{\Psi_{\tTC}}$. Moreover, according to the method for deriving the effective reduced density matrix of PEPS~\cite{Cirac_2011}, we can construct an isometry $\mathcal{V}=(\sigma_L^{-1/2})^{*}{\Psi}_L^{\text{T}}$ applied on physical degrees of freedom in $(R\setminus\partial R)$ such that
\begin{equation}\label{eq_RDM_IGH}
\mathcal{V}\Psi^{*}_L\mathcal{T}_{\pmb{x}}^{\dagger}\sigma_L\mathcal{T}_{\pmb{x}}\Psi_L^{\text{T}}\mathcal{V}^{\dagger}\propto\sqrt{\sigma_L^{\text{T}}}\mathcal{T}_{\pmb{x}}^{\dagger}\sigma_L\mathcal{T}_{\pmb{x}}\sqrt{\sigma^{\text{T}}_L}\sim \mathcal{T}_{\pmb{x}}^{\dagger}\sigma_L\mathcal{T}_{\pmb{x}}\sigma^{\text{T}}_L,\notag
\end{equation}
where we use the relation $\sigma_L=\sigma_L^{\dagger}$ and $\Psi_L^{\dagger}\Psi_L\propto\sigma_L$. So the spectrum of $\rho_{\tGH,\pmb{x}}$ and the spectrum of the operator $\mathcal{T}_{\pmb{x}}^{\dagger}\sigma_L\mathcal{T}_{\pmb{x}}\sigma^{\text{T}}_L$ are equal up to a normalization factor.

This method can be used to extract the ES of the Fradkin-Shenker model from the corresponding PEPS of the TC model in a field. In Figs.~\ref{Fig:app_compare}b and d, we show the ES of the Fradkin-Shenker model calculated by combining the quantum channel with the exponentiated perturbed PEPS and the variational PEPS of the TC model in a field, separately. It can be found that the low energy parts of the ES from two methods are almost the same, and there are some slight differences in the higher parts of the ES. So we can combine the quantum channel and the corresponding exponentiated perturbed PEPS of the TC model in a field to calculate the ES of the Fradkin-Shenker model near the fixed points of gapped phases.

In addition, when considering the trivial MES, we just add a project $P_{+}$ to $\mathcal{T}_{\pmb{x}}^{\dagger}\sigma_L\mathcal{T}_{\pmb{x}}\sigma^{\text{T}}_L$, similar to the case shown in Eq.~\eqref{eq_RDM_TC}. In Appendix~\ref{App: Tensor_network_method_for_distillable_TEE}, we use the operator $\mathcal{T}_{\pmb{x}}^{\dagger}\sigma_L\mathcal{T}_{\pmb{x}}\sigma^{\text{T}}_L$ to extract the total and distillable 1/2-R\'enyi entanglement entropies of the Fradkin-Shenker model.

\section{Spontaneous symmetry breaking of emergent 1-form symmetry is robust against measurement}\label{App:explicit_form_of_emergent_1-form_sym}

This section shows that in the deconfined phase where the 1-form 't Hooft loop symmetry is emergent and breaks spontaneously, the measured states also spontaneously break the emergent 1-form 't Hooft loop symmetry. For this purposes, we derive an expression of the emergent 1-form 't Hooft loop symmetry from the first order perturbed ground state of the TC model in a field. We observe that the first order perturbed ground state can be re-expressed as applying a unitary operator $U$ to the fixed point toric code ground state:
\begin{align}
    \ket{\Psi_{\tTC}}&=[U+O(r^2)]\ket{\tTC}\notag\\
    &=\left[\prod_e\exp(i \frac{h_z}{4}A'_{v \ni e}Y_e-i \frac{h_x}{4}B'_{p \ni e}Y_e)+O(r^2)\right]\ket{\tTC}\notag\\
    &=\left[\prod_e\left( 1+i\frac{h_z}{4} A'_{v \ni e}Y_e-i\frac{h_x}{4}iB'_{p \ni e}Y_e\right)+O(r^2)\right]\ket{\tTC}\notag\\
      &=\left[\prod_e \left( 1+\frac{h_z}{4} Z_e +\frac{h_x}{4}X_e\right)+O(r^2)\right]\ket{\tTC},
\end{align}
where $A_{v\ni e}'=A_{v\ni e}X_e$, $B_{p\ni e}'=B_{p\ni e}Z_e$ and $A_{v\ni e}$ ($B_{p\ni e}$) is a vertex (plaquette) operator containing $X_e$ ($Z_e$). Notice that the local unitary gates $U_e$ in the unitary operator $U=\prod_eU_e$ can be decomposed as
$U_e=U_{e,z}U_{e,x}+O(r^2)$,
where $U_{e,z}=\exp(h_z A_{v\ni e}'Y_e/4)$ and $U_{e,x}=\exp(-h_x B_{p\ni e}'Y_e/4)$ could not commute with each other. However, according to the Trotter decomposition, the error arising from non-commuting gates is $O(r^2)$. Since $[U_{e,x},T_{\hat{C}}]=0$ but $[U_{e,z},T_{\hat{C}}]\neq
0$, the emergent 1-form 't Hooft loop symmetry can be derived from the exact 1-form 't Hooft loop operator $T_{\hat{C}}=\prod_{e\in \hat{C}} X_e$:
\begin{equation}
  \tilde{T}_{\hat{C}}=\prod_{e\in \hat{C}}U_eT_{\hat{C}}\prod_{e\in \hat{C}}U_e^{\dagger}=\prod_{e\in \hat{C}}U_{e,z}T_{\hat{C}}\prod_{e\in \hat{C}}U_{e,z}^{\dagger}.
\end{equation}

Next, we check if the measured states spontaneously break the emergent 1-form 't Hooft loop symmetry. Since $\tilde{T}_{\hat{C}_y}\ket{\Psi_{\tTC}}=UT_{\hat{C}_y}U^{\dagger}U\ket{\tTC}=UT_{\hat{C}_y}\ket{\tTC}$ and we can pull through $T_{\hat{C}_y}$ from the physical level to the virtual level, which becomes $\prod_i X_i$, so the emergent 1-form 't Hooft loop symmetry $\tilde{T}_{\hat{C}_y}$ is equivalent to the virtual $\mathbb{Z}_2$ symmetry of the PEPS. Therefore, we can detect if the emergent 1-form symmetry breaks spontaneously via:
\begin{equation}
\bra{\Psi_{\tTC,\pmb{x}}}\tilde{T}_{\hat{C}_y}\ket{\Psi_{\tTC,\pmb{x}}}=\frac{\Tr(\mathcal{T}^{\dagger}_{\pmb{x}}\sigma_L\prod_iX_i\mathcal{T}_{\pmb{x}} \sigma^{\text{T}}_L)}{\Tr(\mathcal{T}^{\dagger}_{\pmb{x}}\sigma_L\mathcal{T}_{\pmb{x}} \sigma_L^{\text{T}})},
\end{equation}
where $\prod_i X_i\sigma_L\neq\sigma_L$ and $\prod_i X_i\mathcal{T}_{\pmb{x}}=\mathcal{T}_{\pmb{x}}\prod_i X_i\not\propto\mathcal{T}_{\pmb{x}}$ (in contrast, when the 1-form 't Hooft loop symmetry is exact, $\prod_i X_i\mathcal{T}_{\pmb{x}}\propto\mathcal{T}_{\pmb{x}}$).
So $\bra{\Psi_{\tTC,\pmb{x}}}\tilde{T}_{\hat{C}_y}\ket{\Psi_{\tTC,\pmb{x}}}$ should decay exponentially with $N$ to zero, and
 $\Tr\left(\tilde{T}_{\hat{C}_y}e^{-nH_{E,\pmb{x}}}\right)/\Tr e^{-nH_{E,\pmb{x}}}$ also decays exponentially with $N$ to zero $\forall n$. From the derivation in Sec.~\ref{TEE_and_1_form_SSB}, we conclude that the TEE can be extracted from the measured state as well as the distillable EE when the 't Hooft loop symmetry is an emergent symmetry.

\section{Tensor network method for calculating the distillable EE}\label{App: Tensor_network_method_for_distillable_TEE}
\begin{figure}
    \centering
    \includegraphics{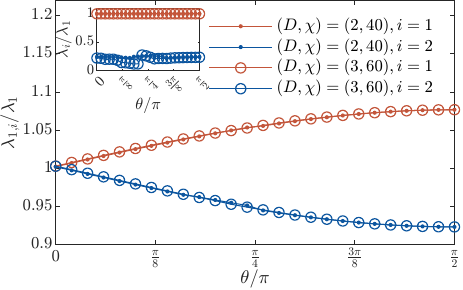}
    \caption{The first two dominant eigenvalues  $\lambda_{1,1}$ and $\lambda_{1,2}$ of $M_1$ defined in Eq.~\eqref{def_M_1} along $\delta=0.15$. Inset: the first two dominant eigenvalues $\lambda_1$ and $\lambda_2$ of $M$ defined in Eq.~\eqref{Trace_rho}. }
    \label{Figure_append_Degeneracy}
\end{figure}

 In this section, we show how to use the tensor network method to calculate the distillable R\'enyi EE. First, as a warm-up, we show how to calculate the total R\'enyi EE~\cite{haller2023quantum}. We consider the the R\'enyi EE of the trivial MES. From the discussion in Sec.~\ref{sec:ES_EH_IGH}, the $1/2$-R\'enyi EE is given by:
\begin{align}
S_{1/2}(\rho_{\tGH})&=2\log\sum_{\pmb{x}}\Tr\left(P_{+}\mathcal{T}_{\pmb{x}}^{\dagger}\sigma_L\mathcal{T}_{\pmb{x}}\sigma^{\text{T}}_L\right)^{1/2}\notag\\
&-\log\sum_{\pmb{x}}\Tr\left(P_{+}\mathcal{T}_{\pmb{x}}^{\dagger}\sigma_L\mathcal{T}_{\pmb{x}}\sigma^{\text{T}}_L\right).
\end{align}
Because we impose the square lattice symmetry to real
PEPS tensors, we have $\sigma_L=\sigma_L^{\text{T}}$ and $\mathcal{T}_{\pmb{x}}=\mathcal{T}_{\pmb{x}}^{\dagger}$. So, further simplification can be made:
\begin{equation}
S_{1/2}=2\log\sum_{\pmb{x}}\Tr\left(P_{+}\sigma_L\mathcal{T}_{\pmb{x}}\right)-\log\sum_{\pmb{x}}\Tr\left(P_{+}\sigma_L\mathcal{T}_{\pmb{x}}\sigma_L \mathcal{T}_{\pmb{x}}\right).\notag
\end{equation}
The two terms in the above equation can be expressed as two tensor networks:
\begin{align}
\sum_{\pmb{x}}\Tr(P_{+}\sigma_L \mathcal{T}_{\pmb{x}})=\frac{1}{2}&
\vcenter{\hbox{
  \includegraphics{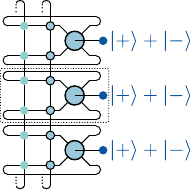}}}=\frac{1}{2}\Tr M_{\text{half}}^N, \notag\\
  \sum_{\pmb{x}}\Tr(P_{+}\sigma_L \mathcal{T}_{\pmb{x}}\sigma_L \mathcal{T}_{\pmb{x}})&=\frac{1}{2}\vcenter{\hbox{
  \includegraphics{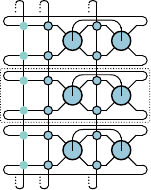}}}=\frac{1}{2}\Tr M^N,\label{Trace_rho}
\end{align}
where the graphic notation of $P_+$ can be found in Fig.~\ref{Figure apppendix_4}, and $M_{\text{half}}$ ($M$) is a transfer matrix shown in the dotted line box in the first (second) equation. So the $1/2$-R\'enyi EE can be expressed as
\begin{equation}
    S_{1/2}=2\log \Tr M_{\text{half}}^N-\log \Tr M^N-\log2.
\end{equation}
When the circumference is infinite, we have $\lim_{N\rightarrow\infty}M^{N}=\lambda\sum_{i=1}^{d}\ket{R_{i}}\bra{L_{i}}$, where $\lambda$, $L_i$($R_i$) and $d$ are the dominant eigenvalue, the $i$-th left (right) dominant eigenvectors and the degeneracy of dominant eigenvalue, respectively, and the dominant eigenvectors satisfy the bi-orthonormal condition $\braket{L_i}{R_j}=\delta_{i,j}$. Similarly, we have $\lim_{N\rightarrow\infty}M_{\text{half}}^{N}=\lambda_{\text{half}}\sum_{i=1}^{d_{\text{half}}}\ket{R_{\text{half},i}}\bra{L_{\text{half},i}}$. So if $N$ is very large, we have
\begin{equation}\label{Renyi_EE_TNS}
 S_{1/2}\approx N \log (\lambda_{\text{half}}^2/\lambda)+\log (d^2_{\text{half}}/d)-\log 2.
\end{equation}
Usually, in the topologically ordered phase, $d_{\text{half}}=d=1$ due to spontaneous 1-form symmetry breaking, and Eq. ~\eqref{Renyi_EE_TNS} is a tensor network version of Eq.~\eqref{REE_of_TC}, so we know the TEE is $\log 2$.

Following the same logic, let us consider the distillable R\'enyi EE of the trivial MES:
\begin{equation}   S_{D,1/2}=\log\sum_{\pmb{x}}\left(\Tr P_{+}\sigma_{L}\mathcal{T}_{\pmb{x}}\right)^2-\log\sum_{\pmb{x}}\Tr\left(P_{+}\sigma_{L}\mathcal{T}_{\pmb{x}}\sigma_{L}\mathcal{T}_{\pmb{x}}\right).
\end{equation}
The graphic notation of the first term is the following tensor network:
\begin{equation}\label{def_M_1}
\sum_{\pmb{x}}\left(\Tr P_{+}\sigma_{L}\mathcal{T}_{\pmb{x}}\right)^2=\frac{1}{4}\vcenter{\hbox{
  \includegraphics{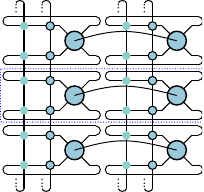}}}=\frac{1}{4}\Tr M_1^N,
\end{equation}
where $M_1$ is the transfer matrix in the vertical direction shown in the dash line box. Therefore, the distillable EE can be expressed as
\begin{equation}
    S_{D,1.2}=\log \Tr M_1^N-\log \Tr M^N-\log 2.
\end{equation}
When the circumference is infinite, we have
$\lim_{N\rightarrow\infty}M_1^{N}=\lambda_1\sum_{i=1}^{d_1}\ket{R_{1,i}}\bra{L_{1,i}}$. So if $N$ is very large, we have
\begin{equation}\label{distillable_TEE}
 S_{D,1/2}\approx N \log (\lambda_1/\lambda)+\log (d_1/d)-\log 2.
\end{equation}
Therefore, the distillable R\'enyi EE also satisfies the area law plus a constant correction $\gamma_D=\log (d_1/d)-\log 2$.

We calculate the first and the second dominant eigenvalues of $M_1$ and $M$ along $\delta=0.15$, as shown in Fig.~\ref{Figure_append_Degeneracy}. It can be found that when $\theta\neq 0$ ($h_z\neq 0$) where the 1-form 't Hooft loop symmetry is emergent, both the dominant eigenvalues of $M$ and $M_1$ are not degenerate, so $d_1=d=1$, and the constant correction is $\gamma_D=\log 2$, according to Eq.~\eqref{distillable_TEE}. When $\theta=0$, where the 1-form 't Hooft loop symmetry is exact, the dominant eigenvalue of $M_1$ becomes 2-fold degenerate, so $d_1=2$ and $d=1$, and the constant correction is $0$, according to Eq.~\eqref{distillable_TEE}. So the TEE can not be extracted from the distillable EE when $\theta=0$. In addition, the dominant eigenvalues $\lambda$ and $\lambda_1$ of $M$ and $M_1$ are very close, so Eq.~\eqref{distillable_TEE} implies that the density of the distillable R\'enyi EE is very small.

\bibliography{sample.bib}

\end{document}